\documentclass[12pt,a4paper]{article}
\usepackage{jheppub}

\usepackage{graphicx}
\usepackage{amsmath,amssymb}

\DeclareMathOperator{\tr}{tr}

\newcommand{\del}{\partial}

\title{Twistors, Harmonics and Holomorphic Chern-Simons}
\author{Burkhard U.~W.~Schwab,}
\author{Cristian Vergu}
\affiliation{ETH Z\"urich, Institut f\"ur theoretische Physik,\\
Wolfgang Pauli Strasse 27,\\
8093 Z\"urich, Switzerland}
\emailAdd{schwabbu@itp.phys.ethz.ch}
\emailAdd{verguc@itp.phys.ethz.ch}

\abstract{We show that the off-shell $\mathcal{N}=3$ action of $\mathcal{N}=4$ super Yang-Mills can be written as a holomorphic Chern-Simons action whose Dolbeault operator $\bar{\partial}$ is constructed from a complex-real (CR) structure of harmonic space.  We also show that the local space-time operators can be written as a Penrose transform on the coset $SU(3)/(U(1) \times U(1))$.  We observe a strong similarity to ambitwistor space constructions.}


\begin{document}
\maketitle

\section{Introduction}
\label{sec:introduction}

In recent years the study of twistor theory has been undergoing a \textit{renaissance}, following work by Witten (see ref.~\cite{Witten:2003nn}).  One of the highlights of these recent developments was the construction in the $\mathcal{N}=4$ super-Yang-Mills theory of a supersymmetric non-Abelian Wilson loop by Mason and Skinner in ref.~\cite{Mason:2010yk} (the space-time version of this supersymmetric Wilson loop was constructed independently by Caron-Huot in ref.~\cite{CaronHuot:2010ek}).

The computation (see refs.~\cite{Mason:2010yk, CaronHuot:2010ek}) of these supersymmetric Wilson loops on polygonal contours, revealed that -- in the planar limit -- they reproduce the planar limit of scattering amplitudes.  This supported earlier observations linking polygonal Wilson loops in $\mathcal{N}=4$ in the planar limit and MHV scattering amplitudes~\cite{Anastasiou:2003kj, Bern:2005iz, Bern:2006vw, arXiv:0707.1153, arXiv:0709.2368, arXiv:0712.1223, arXiv:0712.4138, arXiv:0803.1465, Anastasiou:2009kna, ArkaniHamed:2010kv, Kosower:2010yk}.  Remarkably, the equivalence between scattering amplitudes and Wilson loops is supported by strong coupling computations as well~\cite{arXiv:0705.0303, arXiv:0807.3196}.  In ref.~\cite{Bullimore:2011ni} it was shown using the loop equations satisfied by the supersymmetric Wilson loops that their integrand satisfies the same recursion relation as the loop level extension of the BCFW recursion presented in ref.~\cite{ArkaniHamed:2010kv}.

Another natural class of observables are correlation functions of local gauge-invariant operators.  The $n$-point correlation functions depend on $n$ coordinates $x_{i}$ and in the light-like limit when $x_{i,i+1}^{2} = 0$ they simplify dramatically.  In a series of papers~\cite{Alday:2010zy, Eden:2010zz, Eden:2010ce, Eden:2011yp, Eden:2011ku} Alday, Eden, Korchemsky, Maldacena and Sokatchev have shown that in this light-like limit one can also extract the same information contained in the scattering amplitudes and in the Wilson loops.  An understanding of this new equivalence from the twistor point of view was provided in ref.~\cite{Adamo:2011dq} by considering correlation functions of Konishi operators.

One of the striking features of the correspondence between correlation functions and chiral supersymmetric Wilson loops or scattering amplitudes is that the correlation functions need to be chirally projected in some way. As a consequence a lot of information contained in the correlation functions is lost.  After this projection, part of the manifest symmetry of the correlation functions (like the antichiral $\bar{Q}$ supersymmetry) becomes anomalous.  A proposal for the anomaly of the $\bar{Q}$ operator has been put forward in refs.~\cite{Bullimore:2011kg, CaronHuot:2011kk}.  

We may also try to define nonchiral supersymmetric Wilson loop as it was proposed in ref.~\cite{CaronHuot:2011ky}.  The construction of such a nonchiral supersymmetric Wilson loop was completed and studied in refs.~\cite{Beisert:2012gb, Beisert:2012xx}.  It turns out that nonchiral supersymmetric Wilson loops would most naturally be formulated in ambitwistor space, but ambitwistor theory is poorly understood.
\medskip

We are therefore led to consider other nonchiral formulations of $\mathcal{N}=4$ super-Yang-Mills theory.  One such nonchiral formulation is the $\mathcal{N}=3$ theory of Galperin et al.~\cite{Galperin:1984bu}.  Before we start discussing this theory let us discuss the simpler example of selfdual $\mathcal{N}=4$ super-Yang-Mills.

In ref.~\cite{Sokatchev:1995nj} an off-shell action was written for the selfdual $\mathcal{N}=4$ super-Yang-Mills of Siegel~\cite{Siegel:1992ev}.  The Lagrangian of ref.~\cite{Sokatchev:1995nj} has the form
\begin{equation}
  \tr \left(A^{++} D^{+ \alpha} A_{\alpha}^{+} - \frac 1 2 A^{+ \alpha} D^{++} A_{\alpha}^{+} + A^{++} A^{+ \alpha} A_{\alpha}^{+}\right),
\end{equation} where $A^{++}$ and $A^{+ \alpha}$ are two fields and $D^{+ \alpha}$ and $D^{++}$ are the derivatives.  This action can be written in a better way, by introducing two vielbeine $e^{- \alpha}$ and $e^{--}$ to form a one-form connection
\begin{equation}
  A = e^{--} A^{++} + e^{- \alpha} A_{\alpha}^{+}
\end{equation} and a $\bar{\partial}$ operator
\begin{equation}
  \bar{\partial} = e^{--} D^{++} + e^{- \alpha} D_{\alpha}^{+}.
\end{equation}  With these we can write a three-form Lagrangian
\begin{equation}
  \tr \left(A \bar{\partial} A + \frac 2 3 A \wedge A \wedge A\right),
\end{equation} which reproduces the previous one in components.  This reformulation is implicit in ref.~\cite{Witten:2003nn}, for example (see also ref.~\cite{Boels:2006ir} where the action in terms of component space-time fields was worked out).

In ref.~\cite{Galperin:1984bu}, a similar action was found for the full $\mathcal{N}=4$ theory which exhibits $\mathcal{N}=3$ supersymmetry off-shell.  That action is more complicated.  Its Lagrangian reads
\begin{multline}
  \label{eq:neq3-components}
  \tr\Bigl(A^{(2,-1)} (D^{(-1,2)} A^{(1,1)} - D^{(1,1)} A^{(-1,2)}) -
  A^{(-1,2)} (D^{(2,-1)} A^{(1,1)} - D^{(1,1)} A^{(2,-1)}) +\\
  A^{(1,1)} (D^{(2,-1)} A^{(-1,2)} - D^{(-1,2)} A^{(2,-1)}) - (A^{(1,1)})^{2} - 2 A^{(1,1)} [A^{(2,-1)}, A^{(-1,2)}]\Bigr),
\end{multline} where $A^{(2,-1)}$, $A^{(1,1)}$ and $A^{(-1,2)}$ are three dynamical fields and $D^{(-1,2)}$, $D^{(1,1)}$ and $D^{(2,-1)}$ are three derivatives.  This action has some of the expected features of a Chern-Simons action, like being cubic in the fields and first order in derivatives, but it also contains a slightly puzzling term like $(A^{(1,1)})^{2}$, which is quadratic but does not have any derivatives.

One could solve for the field $A^{(1,1)}$ by using its equations of motion and plugging the solution back in the action, but this would lead to a more complicated action which is quartic in the fields (see ref.~\cite[chap.~12]{Galperin:2001uw}).  Therefore it is preferable to keep $A^{(1,1)}$.

We will show that the origin of the puzzling term $(A^{(1,1)})^{2}$ is in the torsion of the coset $SU(3)/(U(1) \times U(1))$.  Schematically, the reason is as follows.  We can define a connection
\begin{equation}
  A = e^{(-2,1)} A^{(2,-1)} + e^{(-1,-1)} A^{(1,1)} + e^{(1,-2)} A^{(-1,2)},
\end{equation} where $e^{(-2,1)}$, $e^{(-1,-1)}$ and $e^{(1,-2)}$ are one-form vielbeine and we can also define a differential $\bar{\partial}$ by
\begin{equation}
  \bar{\partial} = e^{(-2,1)} D^{(2,-1)} + e^{(-1,-1)} D^{(1,1)} + e^{(1,-2)} D^{(-1,2)}.
\end{equation}

When computing $\bar{\partial} A$, the differential $\bar{\partial}$ can act on the component fields in $A$, but also on the vielbeine $e^{(-2,1)}$, $e^{(-1,-1)}$ and $e^{(1,-2)}$.  As we will show, the torsion makes the action of $\bar{\partial}$ on the vielbeine non-trivial
\begin{equation}
  \bar{\partial} e^{(-1,-1)} = -e^{(-2,1)} \wedge e^{(1,-2)}, \qquad
  \bar{\partial} e^{(1,-2)} = 0, \qquad
  \bar{\partial} e^{(-2,1)} = 0.
\end{equation}  With these preparations, the Lagrangian can be written as a holomorphic Chern-Simons Lagrangian\footnote{In order to obtain the $\mathcal{N}=4$ theory in $(3,1)$ signature a reality condition must be imposed on the connection $A$, as we will discuss below.}
\begin{equation}
    \tr \left(A \bar{\partial} A + \frac 2 3 A \wedge A \wedge A\right).
\end{equation}  When written in components the Lagrangian above matches the one in eq.~\eqref{eq:neq3-components}.  We have not been able to find this formulation anywhere in the literature.

We should note here that even though the $\mathcal{N}=3$ Lagrangian looks the same as the one for the $\mathcal{N}=4$ self-dual theory, the interpretation is different since the one-forms $A$ are defined on a different space and the $\bar{\partial}$ differential is defined in a different way. However, either way, only three gauge fields are necessary to fully describe $\mathcal{N}=4$ super Yang-Mills theory.

Besides its simplicity, the formulation as a holomorphic Chern-Simons theory has the advantage of emphasizing the underlying geometry of the problem which is helpful when understanding symmetries.  As an example, under superconformal transformations the components $A^{(2,-1)}$, $A^{(1,1)}$ and $A^{(-1,2)}$ of the connection $A$ mix among themselves but the one-form $A$ just transforms by a Lie derivative (see sec.~\ref{sec:kill-vect}).  Also, when computing the propagator, it will probably be best to compute $\langle A_{(1)} A_{(2)}\rangle$ instead of two-point functions of component fields.

The paper is organized as follows.  In sec.~\ref{sec:general-philosophy} we present the general philosophy behind the harmonic superspace constructions in the general setting.  In sec.~\ref{sec:neq3-theory} we review the construction of the $\mathcal{N}=3$ theory based on the $SU(3)/(U(1) \times U(1))$ coset and we present its formulation as a holomorphic Chern-Simons theory.  In sec.~\ref{sec:selfdual-theory} we present the self-dual theory, mostly in the language of ref.~\cite{Sokatchev:1995nj} and make contact with the twistor constructions.  In the next section~\ref{sec:super-wilson-loops} we review the construction of super Wilson loops.  These last two sections do not contain anything new, but they cover useful background for the construction of local operators in sec.~\ref{sec:local-operators}.  We end with conclusions and a number of appendices.

\emph{[Note added in version 3: At the time of the formation of this paper, the authors were not aware of the work \cite{Roslyi:1985hn} which anticipated many of the results in sec.~\ref{sec:neq3-theory}.]}

\section{General philosophy}
\label{sec:general-philosophy}

Before we go on to discuss concrete examples, let us describe the basic strategy in the harmonic superspace constructions.  To start, we need to pick a superspace which contains space-time and some odd coordinates.  Then, as usual, we introduce supersymmetry covariant derivatives.  In gauge theories to each of these supersymmetry covariant derivatives corresponds a gauge covariant derivative.  In other words, we introduce gauge connections for each supersymmetry covariant derivative.

From the gauge covariant derivatives we define curvatures, or field strengths, paying attention to subtracting the torsion terms, which appear in superspace.  It is well-known that these field strengths satisfy some constraints.  In cases with maximal (or near maximal) supersymmetry like $\mathcal{N}=4$ or $\mathcal{N}=3$ for four-dimensional Yang-Mills, these constraints are so strong that they imply the equations of motion.  Said differently, when we have a lot of supersymmetry, the action is uniquely determined.\footnote{Up to terms which are supersymmetric by themselves and which do not contribute to the equations of motion, like $\int \tr (F \wedge F)$ for Yang-Mills theory.}  Then we can work out the action of supersymmetry on fields, which turns out to be nonlinear.  Moreover, in the most interesting cases the supersymmetry algebra only closes on-shell which makes it difficult to work out the consequences of supersymmetry.

The harmonic superspace approach attempts to introduce auxiliary fields such that (at least some) supersymmetry is realized linearly and off-shell.  The construction proceeds by introducing extra bosonic coordinates $u$, which parametrize a coset $G/H$.  In fact, space-time and superspace can themselves be written as cosets.  Standard methods allow us to compute covariant derivatives on $G/H$.  Then, it is sensible to think of the theory as living on a bigger space so the gauge connections depend on the extra variables $u$.

The next step is to make a change of coordinates on this bigger space such that the constraints discussed above take a simpler form.  From the covariant derivatives defined beforehand, we need to pick an integrable distribution which implies the constraints.  A distribution is generated by a set of vector fields (or derivation operators) whose commutator is expressible in terms of themselves.  The distributions we will work with contain both Grassmann even and Grassmann odd elements.  In favorable cases the constraints imposed by the Grassmann odd elements of the distribution can be solved by restricting the dependence of the connection on the odd coordinates.  The simplest example of this phenomenon is the case of chiral fields where the constraints are $\bar{D} \phi = 0$.  After solving these easy constraints, we are still left with some more.  These remaining constraints will be interpreted as the equations of motion of a new action.

The integrable distribution mentioned above generates a CR\footnote{CR stands for Cauchy-Riemann or Complex-Real, according to taste.} structure.  CR structures are central in harmonic or twistor constructions.  We present a short discussion of CR structure and work out some explicit examples in sec.~\ref{sec:cr-manifolds}.

The strategy we have presented can be turned around in the following way.  We take as a starting point a symmetry group $G$, for example $SU(2,2\vert 3)$.  Then, we look for a manifold $M$ with a CR structure defined by an integrable distribution $L$ of even rank three and arbitrary odd rank $\kappa$.  The manifold $M$ and the CR structure should be such that the group of diffeomorphisms of $M$ which preserve the CR structure is isomorphic to the symmetry group $G$.

Finding a manifold $M$ with a CR structure which is preserved by $G$ is not necessarily straightforward, but it can be suggested by the usual analysis of constraints.  Using this data we build a field theory of a connection $A$ on $M$ which is holomorphic Chern-Simons, and which has the right symmetries.\footnote{The only issue that can arise is in defining the integration measure for the Lagrangian in a way which preserves the symmetries, but usually this does not pose a problem.}  In interesting cases like $\mathcal{N}=4$ super-Yang-Mills the symmetry group determines the theory completely.  Then we are left with the challenge of writing the local space-time operators in terms of the connection $A$.  They will necessarily depend on $A$ in a nonlocal way.  Ensuring the correct gauge transformations is a useful constraint in this construction.

\section{\texorpdfstring{$\mathcal{N}=3$}{N=3} theory as holomorphic Chern-Simons}
\label{sec:neq3-theory}

The presentation in this section is, up to a point, heavily inspired by the book~\cite{Galperin:2001uw} by Galperin et al.\ whose conventions we adopt.  The original construction of off-shell $\mathcal{N}=3$ action was done in ref.~\cite{Galperin:1984bu}, but with slightly different notations than in~\cite{Galperin:2001uw}.

The $\mathcal{N}=3$ theory in four dimensions is defined on a superspace with coordinates $z=(x, \theta, \bar{\theta})$, where the odd coordinates $\theta_{i}^{\alpha}$ transform as $\mathbf{3}$ under $SU(3)$ $R$ symmetry group and $\bar{\theta}^{\dot{\alpha} i}$ transform as a $\bar{\mathbf{3}}$.  The supersymmetry covariant derivatives are denoted by $D_{\alpha}^{i}$, $D_{\dot{\alpha} i}$ and $D_{\alpha \dot{\alpha}}$, with an algebra
\begin{equation}
  \left\lbrace D_{\alpha}^{i}, D_{\beta}^{j}\right\rbrace = 0, \qquad
  \left\lbrace D_{\dot{\alpha} i}, D_{\dot{\beta} j}\right\rbrace = 0, \qquad
  \left\lbrace D_{\alpha}^{i}, \bar{D}_{\dot{\alpha} j}\right\rbrace = -2 i \delta_{j}^{i} D_{\alpha \dot{\alpha}}.
\end{equation}

For each supersymmetry covariant derivative we introduce a connection and we define supersymmetry and gauge covariant derivatives $\mathcal{D} = D + A$.  On these derivatives we impose the constraints
\begin{equation}
  \left\lbrace \mathcal{D}_{\alpha}^{i}, \mathcal{D}_{\beta}^{j}\right\rbrace = \epsilon_{\alpha \beta} \bar{W}^{i j}, \qquad
  \left\lbrace \mathcal{D}_{\dot{\alpha} i}, \mathcal{D}_{\dot{\beta} j}\right\rbrace = \epsilon_{\dot{\alpha} \dot{\beta}} W_{i j}, \qquad
  \left\lbrace \mathcal{D}_{\alpha}^{i}, \bar{\mathcal{D}}_{\dot{\alpha} j}\right\rbrace = -2 i \delta_{j}^{i} \mathcal{D}_{\alpha \dot{\alpha}}.
\end{equation}  We have the following reality conditions $(\mathcal{D}_{\alpha}^{i})^{\dagger} = \bar{\mathcal{D}}_{\dot{\alpha} i}$, $(\mathcal{D}_{\alpha \dot{\alpha}})^{\dagger} = - \mathcal{D}_{\alpha \dot{\alpha}}$ and $(W_{ij})^{\dagger} = \bar{W}^{ij}$.  The superfield $\bar{W}^{ij}$ is antisymmetric, $\bar{W}^{ij} = -\bar{W}^{ji}$ and by the reality conditions it is the only independent curvature. It is well-known that these constraints describe $\mathcal{N}=3$ super Yang-Mills theory.

These constraints can be rewritten in an equivalent way by introducing two $SU(3)$ triplets: $\xi_{i}$ transforming in the $\mathbf{3}$ of $SU(3)$ and $\eta^{i}$ transforming in the $\bar{\mathbf{3}}$ of $SU(3)$ and such that $\xi_{i} \eta^{i} = 0$.  Using these variables, we can define $\mathcal{D}_{\alpha} = \xi_{i} \mathcal{D}_{\alpha}^{i}$ and $\bar{\mathcal{D}}_{\bar{\alpha}} = \eta^{i} \bar{\mathcal{D}}_{\dot{\alpha} i}$.  The constraints then become
\begin{equation}
  \left\lbrace \mathcal{D}_{\alpha}, \mathcal{D}_{\beta}\right\rbrace = 0,\quad
  \left\lbrace \bar{\mathcal{D}}_{\dot{\alpha}}, \bar{\mathcal{D}}_{\dot{\beta}}\right\rbrace = 0,\quad
  \left\lbrace \mathcal{D}_{\alpha}, \bar{\mathcal{D}}_{\dot{\beta}}\right\rbrace = 0.
\end{equation}

The triplets $\xi$ and $\eta$ are defined up to a rescaling so $(\xi, \eta) \in Q \subset \mathbb{CP}^{2} \times \mathbb{CP}^{2}$, where $Q$ is a quadric defined by $\xi_{i} \eta^{i} = 0$.  The space $Q$ is six (real) dimensional.  We note here that this is very similar to ambitwistor space, which is defined as a quadric in $\mathbb{CP}^{3} \times \mathbb{CP}^{3}$.

We can also show that $\xi$ and $\eta$ parametrize a coset $SU(3)/(U(1) \times U(1))$.  If we start with $\xi$ and $\eta$, we form a $3 \times 3$ unitary matrix $u$ of unit determinant,
\begin{equation}
  u = \left(\frac{\xi}{\lvert \xi\rvert}, \frac{\bar{\eta}}{\lvert \eta\rvert}, \frac {\bar{\xi} \times \eta}{\lvert \xi\rvert \lvert \eta\rvert}\right),
\end{equation} where $\lvert \xi\rvert^{2} = \xi \cdot \bar{\xi}$ and similarly for $\eta$.  However, if we take $\xi, \eta \in \mathbb{CP}^{2}$, then we see that under $\xi \to \xi e^{i \phi_{1}}$, $\eta \to \eta e^{i \phi_{2}}$, the matrix $u$ is not invariant.  In order to obtain invariance we need to identify the matrices obtained by these rescalings, which can be achieved by taking the coset by this $U(1) \times U(1)$ group.

We can decide to work with the variables $(\xi, \eta)$ or work with the matrix $u \in SU(3)$.  In the following we will work with the matrix $u$.  We denote the matrix elements by $u_{i}^{I}$, with $i = 1,2,3$ and $I$ is labeled by the $U(1) \times U(1)$ charges $I = (1,0), (0,-1), (-1,1)$.   Sometimes it is convenient to use a shorter notation where $I$ range over $1,2,3$, with the understanding that these labels correspond to the charges $(1,0), (0,-1), (-1,1)$.  Since $u \in SU(3)$, we also have $(u_{i}^{I})^{*} = u_{I}^{i}$ and $\det u_{i}^{I} = 1$.  Using $u$'s instead of $(\xi, \eta)$, the constraints can be written
\begin{equation}
  \left\lbrace \mathcal{D}_{\alpha}^{(1,0)}, \mathcal{D}_{\beta}^{(1,0)}\right\rbrace = 0, \quad
  \left\lbrace \bar{\mathcal{D}}_{\dot{\alpha}}^{(0,1)}, \bar{\mathcal{D}}_{\dot{\beta}}^{(0,1)}\right\rbrace = 0, \quad
  \left\lbrace \mathcal{D}_{\alpha}^{(1,0)}, \bar{\mathcal{D}}_{\dot{\alpha}}^{(0,1)}\right\rbrace = 0.
\end{equation}

We can now define covariant derivatives on the coset $SU(3)/(U(1) \times U(1))$.  They are computed in detail in sec.~\ref{sec:coset-spaces} so we will only list them here
\begin{alignat}{3}
  D^{(-2,1)} &= u_{i}^{3} \frac {\partial}{\partial u_{i}^{1}}, &\qquad
  D^{(-1,-1)} &= u_{i}^{2} \frac {\partial}{\partial u_{i}^{1}}, &\qquad
  D^{(-1,2)} &= u_{i}^{3} \frac {\partial}{\partial u_{i}^{2}},\\
  D^{(1,-2)} &= u_{i}^{2} \frac {\partial}{\partial u_{i}^{3}}, &\qquad
  D^{(1,1)} &= u_{i}^{1} \frac {\partial}{\partial u_{i}^{2}}, &\qquad
  D^{(2,-1)} &= u_{i}^{1} \frac {\partial}{\partial u_{i}^{3}}.
\end{alignat}

\begin{figure}
  \centering
  \includegraphics{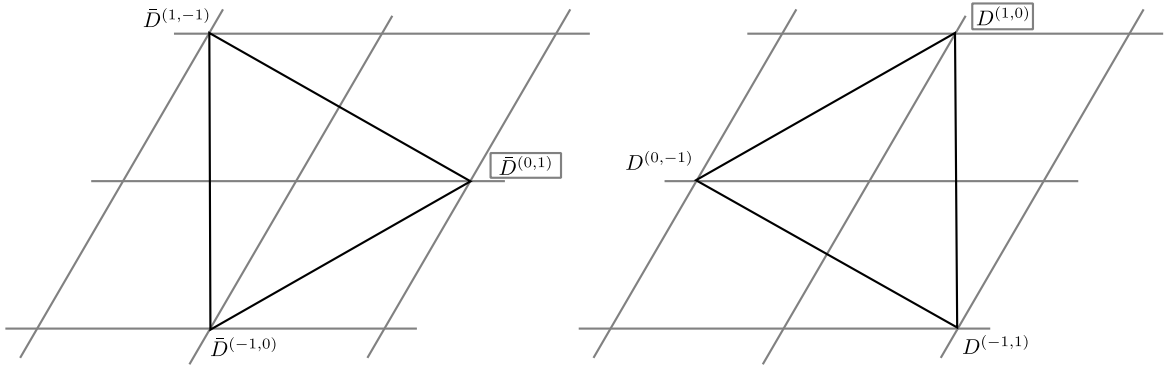}
  \caption{The fermionic derivatives and their $U(1) \times U(1)$ charges.  The boxed generators are the odd generators of the distribution.}
  \label{fig:su3f}
\end{figure}

The algebra of the superspace covariant derivatives $\mathcal{D}_{\alpha}^{(1,0)}$, $\bar{\mathcal{D}}_{\dot{\alpha}}^{(0,1)}$ and the $SU(3)/(U(1) \times U(1))$ covariant derivatives listed above can be easily computed.  We only list the ones which are relevant for us later
\begin{subequations}
  \label{eq:su3-harmonic-super}
\begin{alignat}{3}
  [D^{(2,-1)}, \mathcal{D}_{\alpha}^{(1,0)}] &= 0, \quad
  [D^{(-1,2)}, \mathcal{D}_{\alpha}^{(1,0)}] &= 0, \quad
  [D^{(1,1)}, \mathcal{D}_{\alpha}^{(1,0)}] &= 0,\\
  [D^{(2,-1)}, \bar{\mathcal{D}}_{\dot{\alpha}}^{(0,1)}] &= 0, \quad
  [D^{(-1,2)}, \bar{\mathcal{D}}_{\dot{\alpha}}^{(0,1)}] &= 0, \quad
  [D^{(1,1)}, \bar{\mathcal{D}}_{\dot{\alpha}}^{(0,1)}] &= 0.
\end{alignat}
\end{subequations}

Up to now all the gauge connections were independent on the variables $u$.  We can introduce gauge fields $A^{(q_{1}, q_{2})}$ and construct gauge covariant derivatives $\mathcal{D}^{(q_{1}, q_{2})} = D^{(q_{1}, q_{2})} + A^{(q_{1}, q_{2})}$.  We will also allow the gauge connections to depend on $u$.  Of course, the connections $A^{(q_{1}, q_{2})}$ are flat so the algebra of gauge covariant derivatives $\mathcal{D}^{(q_{1}, q_{2})}$ is the same as the algebra of covariant derivatives $D^{(q_{1}, q_{2})}$ (see eq.~\eqref{eq:su3-cov-algebra})
\begin{alignat}{2}
  [\mathcal{D}^{(-2,1)}, \mathcal{D}^{(1,-2)}] &= \mathcal{D}^{(-1,-1)}, &\quad
  [\mathcal{D}^{(-1,-1)}, \mathcal{D}^{(-1,2)}] &= \mathcal{D}^{(-2,1)},\\
  [\mathcal{D}^{(1,1)}, \mathcal{D}^{(-2,1)}] &= \mathcal{D}^{(-1,2)}, &\quad
  [\mathcal{D}^{(-1,2)}, \mathcal{D}^{(2,-1)}] &= \mathcal{D}^{(1,1)},\\
  [\mathcal{D}^{(1,-2)}, \mathcal{D}^{(1,1)}] &= \mathcal{D}^{(2,-1)}, &\quad
  [\mathcal{D}^{(2,-1)}, \mathcal{D}^{(-1,-1)}] &= \mathcal{D}^{(1,-2)}.
\end{alignat}  After covariantizing the harmonic derivatives, we should replace them in eqs.~\eqref{eq:su3-harmonic-super} by their covariant versions.

Now we pick an integrable distribution\footnote{This choice is not unique. However, it doesn't seem to be possible to choose a basis of commuting vectors for this distribution.} generated by $\mathcal{D}_{\alpha}^{(1,0)}$, $\bar{\mathcal{D}}_{\dot{\alpha}}^{(0,1)}$, $\mathcal{D}^{(2,-1)}$, $\mathcal{D}^{(-1,2)}$ and $\mathcal{D}^{(1,1)}$.  All the (anti)-commutators of these derivatives are zero with the exception of $[\mathcal{D}^{(-1,2)}, \mathcal{D}^{(2,-1)}] = \mathcal{D}^{(1,1)}$.  As we mentioned before, the constraints involving the fermionic covariant derivatives can be solved by going to a gauge where their connections vanish (which is possible because they are flat).  Then, the constraints involving even and odd derivatives can be solved by taking the connections $A^{(2,-1)}$, $A^{(-1,2)}$ and $A^{(1,1)}$ to depend on a restricted set of variables $z = \left(x_{A}, \theta_{\alpha}^{(1,-1)}, \theta_{\alpha}^{(0,1)}, \bar{\theta}_{\dot{\alpha}}^{(1,0)}, \bar{\theta}_{\dot{\alpha}}^{(-1,1)}, u\right)$, where $x_{A}$ is such that $D_{\alpha}^{(1,0)} x_{A}^{\beta \dot{\beta}} = 0$ and $\bar{D}_{\dot{\alpha}}^{(0,1)} x_{A}^{\beta \dot{\beta}} = 0$.  Finally we are left with three constraints, arising from the harmonic derivatives.

These constraints can be written explicitly and an action from which they follow as equations of motion can be found.  However, and this is where our approach differs from the usual treatment, in order to get the holomorphic Chern-Simons action, we will think of these connections as components of a differential one-form, defined as
\begin{equation}
  \label{eq:connection-su3}
  A = e^{(-2,1)} A^{(2,-1)} + e^{(1,-2)} A^{(-1,2)} + e^{(-1,-1)} A^{(1,1)},
\end{equation} where $e^{(-2,1)}$, $e^{(1,-2)}$ and $e^{(-1,-1)}$ are vielbeine dual to the covariant derivatives $D^{(2,-1)}$, $D^{(-1,2)}$ and $D^{(1,1)}$, respectively.  They are computed in sec.~\ref{sec:coset-spaces}, but we also list them here for convenience:
\begin{equation}
  e^{(-2,1)} = u_{1}^{i} d u_{i}^{3}, \qquad
  e^{(1,-2)} = u_{3}^{i} d u_{i}^{2}, \qquad
  e^{(-1,-1)} = u_{1}^{i} d u_{i}^{2}.
\end{equation}

We call the $p$-forms which can be decomposed only on $e^{(-2,1)}$, $e^{(1,-2)}$ and $e^{(-1,-1)}$, $(0,p)$ forms.  For example, the connection $A$ defined in eq.~\eqref{eq:connection-su3} is a $(0,1)$ form.

The next ingredient we need is a Dolbeault operator $\bar{\partial}$.  For a CR manifold there is a standard construction of a Dolbeault operator, described in sec.~\ref{sec:cr-manifolds}.  When acting on $(0,p)$ forms,\footnote{This also includes functions, which are $(0,0)$ forms.} the Dolbeault operator $\bar{\partial}$ can be taken to be
\begin{equation}
  \label{eq:dolbeault-su3}
  \bar{\partial} = e^{(-2,1)} D^{(2,-1)} + e^{(1,-2)} D^{(-1,2)} + e^{(-1,-1)} D^{(1,1)}.
\end{equation}  The action on general $(p,q)$ forms is slightly more involved but fortunately we will not need it.  The presence of torsion makes the action of $\bar{\partial}$ a bit unusual
\begin{equation}
  \bar{\partial} e^{(-1,-1)} = -e^{(-2,1)} \wedge e^{(1,-2)}, \qquad
  \bar{\partial} e^{(1,-2)} = 0, \qquad
  \bar{\partial} e^{(-2,1)} = 0.
\end{equation}

Now we can define a field strength $(0,2)$ form $F = \bar{\partial} A + A \wedge A$.  In components, this reads
\begin{equation}
\begin{aligned}
  F &= e^{(-2,1)} \wedge e^{(1,-2)} \left(D^{(2,-1)} A^{(-1,2)} - D^{(-1,2)} A^{(2,-1)} + [A^{(2,-1)}, A^{(-1,2)}] - A^{(1,1)}\right) +\\
  &e^{(-2,1)} \wedge e^{(-1,-1)} \left(D^{(2,-1)} A^{(1,1)} - D^{(1,1)} A^{(2,-1)} + [A^{(2,-1)}, A^{(1,1)}]\right) +\\
  &e^{(1,-2)} \wedge e^{(-1,-1)} \left(D^{(-1,2)} A^{(1,1)} - D^{(1,1)} A^{(-1,2)} + [A^{(-1,2)}, A^{(1,1)}]\right).
\end{aligned}
\end{equation}  The components of $F$ are exactly the remaining constraints and we see that they can be interpreted as an equation of motion for $A$, $F = \bar{\partial} A + A \wedge A = 0$.

The equation of motion setting a connection to be flat $F=0$ arises naturally from a Chern-Simons action with Lagrangian $\tr\left(A \bar{\partial} A + \tfrac 2 3 A \wedge A \wedge A\right)$.  Keeping in mind that this Chern-Simons Lagrangian is a $(0,3)$ form, what should we integrate over to get the action?  The answer is to introduce a ``form'' $\Omega$, defined as
\begin{equation}
  \Omega = d^{4} x_{A}d^{8}\theta\; e^{(1,1)} \wedge e^{(2,-1)} \wedge e^{(-1,2)}.
\end{equation}  which we can use to write the action in $\mathcal{N}=3$ harmonic superspace as
\begin{equation}
  \label{eq:neq3-harmonic-action}
  S[A] = \int \Omega \wedge \tr\left(A \bar{\partial} A + \frac 2 3 A \wedge A \wedge A\right),
\end{equation} Here $d^{8} \theta = d^{2} \theta^{(1,-1)} d^{2} \theta^{(0,1)} d^{2} \bar{\theta}^{(1,0)} d^{2} \bar{\theta}^{(-1,1)}$.  Notice that the $U(1) \times U(1)$ weights cancel between $e^{(1,1)} \wedge e^{(2,-1)} \wedge e^{(-1,2)}$ which has weights $(2,2)$ and $d^{8} \theta$ which has weights $(-2,-2)$. 

Several comments are in order.  First, the fermionic coordinates need to be integrated since the notion of differential forms does not really apply to them.  After the fermionic integration we are left with an integral over a four-dimensional contour\footnote{It is easy to see that the space coordinates $x_{A}$ are not real.} in $\mathbb{C}^{4}$ parametrized by $x_{A}$, times $Q \subset \mathbb{CP}^{2} \times \mathbb{CP}^{2}$.  Recall that $Q = \lbrace (\xi, \eta) \in \mathbb{CP}^{2} \times \mathbb{CP}^{2} \vert\; \xi \cdot \eta = 0\rbrace$.

The action~\eqref{eq:neq3-harmonic-action} is of holomorphic Chern-Simons type (see ref.~\cite{Witten:1992fb} for the original definition and ref.~\cite{Witten:2003nn} for its version in twistor space).  Holomorphic Chern-Simons is a bit of a misnomer in this case, since what is relevant here is a CR structure, not a complex structure.

In order to obtain the theory in $(3,1)$ signature we need to impose a parity constraint.  The theory has to be invariant under the exchange of $\mathcal{D}_{\alpha}$ and $\bar{\mathcal{D}}_{\dot{\alpha}}$.  We denote the idempotent operation which performs this exchange by $\widetilde{\hspace{5mm}}$.  This implies that
\begin{equation}
  \widetilde{\mathcal{D}_{\alpha}} = \bar{\mathcal{D}}_{\dot{\alpha}}, \qquad
  \widetilde{\xi_{i}} = \eta^{i}, \qquad
  \widetilde{\mathcal{D}_{\alpha}^{i}} = \bar{\mathcal{D}}_{i \dot{\alpha}}.
\end{equation}  Said differently, the operation $\widetilde{\hspace{5mm}}$ swaps the two $\mathbb{CP}^{2}$ and performs complex conjugation on $(x, \theta, \bar{\theta})$ coordinates.  In terms of $u$ coordinates we have
\begin{equation}
  \widetilde{u_{i}^{(1,0)}} = \overline{u_{i}^{(0,-1)}} = u^{i (0,1)}, \qquad
  \widetilde{u_{i}^{(0,1)}} = \overline{u_{i}^{(1,0)}} = u^{i (-1,0)}, \qquad
  \widetilde{u_{i}^{(-1,1)}} = - u^{i (1,-1)}.
\end{equation}

We can show that the reality condition under $\widetilde{\hspace{5mm}}$ operation is $\widetilde{A} = A$.  The integration measure and $\bar{\partial}$ are invariant.

The gauge connection $A$ has a gauge transformation given by $(\bar{\partial} + A) \to g^{-1} (\bar{\partial} + A) g$, where $g$ is an element of the gauge group.  The usual Chern-Simons action is not gauge invariant, but under gauge transformations which are not homotopic to identity it acquires an additive factor.  In order for $e^{i S}$ to be invariant under these gauge transformations the global coefficient of the Chern-Simons theory should be quantized.  We don't know if there are such disconnected gauge transformations in the case we analyzed above, and whether they produce an additive term in the transformation of the action which would necessitate a quantization of the Chern-Simons level.

What are the symmetries of this CR Chern-Simons theory?  The usual Chern-Simons theory is invariant under orientation preserving diffeomorphisms.  For holomorphic Chern-Simons we also need to impose the constraint that the transformations preserve the complex structure.  Finally, for the CR Chern-Simons we need to restrict to the orientation preserving diffeomorphisms which also preserve the distribution which defines the Dolbeault operator.  It is worth noting that if we write the action in terms of component fields $A^{(1,-2)}$, $A^{(2,-1)}$ and $A^{(1,1)}$ the symmetry algebra is much harder to guess.

\section{Selfdual theory}
\label{sec:selfdual-theory}

In this section we present the discussion of the selfdual $\mathcal{N}=4$ super-Yang-Mills theory, using the language of ref.~\cite{Sokatchev:1995nj} and the philosophy of sec.~\ref{sec:general-philosophy}.  Our discussion does not contain anything new, but we feel it is important to review it and contrast it with the features of the $SU(3)/(U(1) \times U(1))$ formulation.

While discussing the selfdual theory we will stay in $(2,2)$ with Lorentz group $SO(2,2)$ or Euclidean signature with Lorentz group $SO(4)$.  The group $SO(2,2)$ is locally isomorphic to $SL(2)_{L} \times SL(2)_{R}$ while $SO(4)$ is locally isomorphic to $SU(2)_{L} \times SU(2)_{R}$.  The spinors transforming under $SL(2)_{L}$ or $SU(2)_{L}$ are indexed by Greek letters from the beginning of the alphabet while the spinors transforming under $SL(2)_{R}$ or $SU(2)_{R}$ are indexed by primed Greek letters from the beginning of the alphabet.

Full superspace has coordinates $(x^{\alpha \alpha'}, \theta_{a}^{\alpha}, \theta^{\alpha' a})$.  The SUSY covariant derivatives are given by
\begin{align}
  D_{\alpha}^{a} &= \frac \partial {\partial \theta_{a}^{\alpha}} +   \frac 1 2 \theta^{\alpha' a} \partial_{\alpha \alpha'},\\
  D_{\beta' b} &= \frac \partial {\partial \theta^{\beta' b}} + \frac   1 2 \theta_{b}^{\beta} \partial_{\beta \beta'},\\
  \partial_{\alpha \alpha'} &= \frac \partial {\partial x^{\alpha \alpha'}}
\end{align} and satisfy the algebra
\begin{equation}
  \lbrace D_{\alpha}^{a}, D_{\beta}^{b}\rbrace = 0, \qquad
  \lbrace D_{\alpha' a}, D_{\beta' b}\rbrace = 0, \qquad
  \lbrace D_{\alpha}^{a}, D_{\beta' b}\rbrace = \delta_{b}^{a} \partial_{\alpha \beta'}.
\end{equation}

To these SUSY covariant derivatives we can associate dual one-forms (or supervielbeine), which are given by
\begin{align}
  e^{\alpha \alpha'} &= d x^{\alpha \alpha'} - \frac 1 2 d \theta_{a}^{\alpha} \theta^{\alpha' a} - \frac 1 2 d \theta^{\alpha' a} \theta_{a}^{\alpha},\\
  e_{a}^{\alpha} &= d \theta_{a}^{\alpha},\\
  e^{\beta' b} &= d \theta^{\beta' b}.
\end{align}  The total differential can be written as
\begin{align}
  d &= d x^{\alpha \alpha'} \frac \partial {\partial x^{\alpha \alpha'}} + d \theta_{a}^{\alpha} \frac \partial {\partial \theta_{a}^{\alpha}} + d \theta^{\beta' b} \frac \partial {\partial \theta^{\beta' b}}\\
  &= e^{\alpha \alpha'} \partial_{\alpha \alpha'} + e_{a}^{\alpha} D_{\alpha}^{a} + e^{\beta' b} D_{\beta' b}.
\end{align}  The differentials of these vielbeine can be written as
\begin{equation}
  d e^{\alpha \alpha'} = e_{a}^{\alpha} \wedge e^{\alpha' a}, \qquad
  d e_{a}^{\alpha} = 0, \qquad
  d e^{\beta' b} = 0.
\end{equation}

Besides the coordinates $(x^{\alpha \alpha'}, \theta_{a}^{\alpha}, \theta^{\alpha' a})$ we will also use harmonic variables, which parametrize a coset $SU(2)/U(1)$.  A matrix $M \in SU(2)$ has elements
\begin{equation}
  M = \begin{pmatrix}
    u^{+ 1} & u^{- 1}\\u^{+ 2} & u^{- 2}
  \end{pmatrix},
\end{equation} where $\pm$ marks the charges under a $U(1)$ subgroup.  In order to have $M \in SU(2)$ we need to take $u^{\pm}_{\alpha'} = (u^{\mp \alpha'})^{*}$ and $u^{+ \alpha'} u_{\alpha'}^{-} = 1$, where $u^{\pm}_{\alpha'} = \epsilon_{\alpha' \beta'} u^{\pm \beta'}$ and $\epsilon$ is the antisymmetric tensor with $\epsilon_{1 2} = -\epsilon_{2 1} = 1$.  Notice that we have taken the columns of $M$ to transform as doublets of $SU(2)_{R}$.

The standard way to compute vielbeine on a coset is to first compute $M^{-1} d M$, which in this case belongs to the $\mathfrak{sl}(2)_{R}$ algebra and then identify the parts which belong to $\mathfrak{u}(1)$ and its complement.  Doing this we find the $SU(2)$ vielbeine
\begin{align}
  e^{--} &= u_{\alpha'}^{-} d u^{- \alpha'},\\
  e^{++} &= -u_{\alpha'}^{+} d u^{+ \alpha'},\\
  \omega^{0} &= -u_{\alpha'}^{-} d u^{+ \alpha'} = -u_{\alpha'}^{+} d u^{- \alpha'}.
\end{align}

These vielbeine have dual covariant derivatives.  Since we are interested in the coset only, we will not need to use the covariant derivative dual to $\omega^{0}$.
\begin{equation}
  e^{--} \leftrightarrow D^{++} = u^{+ \alpha'} \frac \partial {\partial u^{- \alpha'}}, \qquad
  e^{++} \leftrightarrow D^{--} = u^{- \alpha'} \frac \partial {\partial u^{+ \alpha'}}.
\end{equation}

It turns out to be convenient to change coordinates from $(x^{\alpha \alpha'}, \theta_{a}^{\alpha}, \theta^{\alpha' a}, u^{\pm \alpha'})$ to a different set of coordinates.  If we define the antichiral version of $x$ to be $x_{R}^{\alpha \alpha'} = x^{\alpha \alpha'} - \frac 1 2 \theta_{a}^{\alpha} \theta^{\alpha' a}$, then the new coordinates are given by $(x^{\pm \alpha} \equiv x_{R}^{\alpha \alpha'} u_{\alpha'}^{\pm}, \theta^{\pm a} = \theta^{\alpha' a} u_{\alpha'}^{\pm}, \theta_{a}^{\alpha}, u^{\pm \alpha'})$.

The SUSY covariant derivatives in these new coordinates read
\begin{align}
  D_{\alpha}^{a} &= \frac \partial {\partial \theta_{a}^{\alpha}}, \qquad
  D_{\alpha}^{\pm} = \frac \partial {\partial x^{\alpha \mp}},\\
  D_{\beta' b} &= u_{\beta'}^{+} \underbrace{\left(\frac \partial {\partial \theta^{+ b}} + \theta_{b}^{\alpha} \frac \partial {\partial x^{+ \alpha}}\right)}_{D_{b}^{-}} + u_{\beta'}^{-} \underbrace{\left(\frac \partial {\partial \theta^{- b}} + \theta_{b}^{\alpha} \frac \partial {\partial x^{- \alpha}}\right)}_{D_{b}^{+}},\\
  D^{\pm \pm} &= u^{\pm \alpha'} \frac \partial {\partial u^{\mp \alpha'}} + x^{\pm \alpha} \frac \partial {\partial x^{\mp \alpha}} + \theta^{\pm a} \frac \partial {\partial \theta^{\mp a}}.
\end{align}

The dual vielbeine to the SUSY covariant derivatives in the new coordinates are
\begin{align}
  D_{\alpha}^{a} &\leftrightarrow e_{a}^{\alpha} = d \theta_{a}^{\alpha},\\
  D_{b}^{-} &\leftrightarrow e^{+ b} = D \theta^{+ b} + \theta^{- b}   u_{\alpha'}^{+} d u^{+ \alpha'},\\
  D_{b}^{+} &\leftrightarrow e^{- b} = D \theta^{- b} - \theta^{+ b}   u_{\alpha'}^{-} d u^{- \alpha'},\\
  D_{\alpha}^{-} &\leftrightarrow e^{+ \alpha} = D x^{+ \alpha} + \theta_{b}^{\alpha} D \theta^{+ b} + (x^{- \alpha} - \theta^{- a} \theta_{a}^{\alpha}) u_{\alpha'}^{+} d u^{+ \alpha'},\\
  D_{\alpha}^{+} &\leftrightarrow e^{- \alpha} = D x^{- \alpha} + \theta_{b}^{\alpha} D \theta^{- b} - (x^{+ \alpha} - \theta^{+ a} \theta_{a}^{\alpha}) u_{\alpha'}^{-} d u^{- \alpha'},\\
  D^{++} &\leftrightarrow e^{--} = u_{\alpha'}^{-} d u^{- \alpha'},\\
  D^{--} &\leftrightarrow e^{++} = -u_{\alpha'}^{+} d u^{+ \alpha'},
\end{align} where
\begin{equation}
  D x^{\pm \alpha} = d x^{\pm \alpha} \pm \omega^{0} x^{\pm \alpha}, \qquad
   D \theta^{\pm a} = d \theta^{\pm a} \pm \omega^{0} \theta^{\pm a}.
\end{equation}

The transformation from the old vielbeine to the new ones is given by
\begin{equation}
  e^{\alpha' a} = u^{+ \alpha'} e^{- a} - u^{- \alpha'} e^{+ a}, \qquad
  e^{\alpha \alpha'} = u^{+ \alpha'} e^{- \alpha} - u^{- \alpha'} e^{+ \alpha}.
\end{equation}

Let us now formulate the constraints defining the selfdual theory.  Our approach will be similar to the construction of ref.~\cite{Sokatchev:1988qr}, but will differ from it in some details.  In ref.~\cite{Sokatchev:1988qr} the theory was not formulated in terms of differential forms on superspace, as we will do below.  We start in $\mathcal{N}=4$ antichiral superspace.  The chiral coordinate $\theta_{a}^{\alpha}$ will not appear explicitly in the rest of the analysis and it can be thought of as taking some fixed value.  Also, we will not use the chiral derivative $D_{\alpha}^{a}$ at all.

It can be shown that the right constraints defining the $\mathcal{N}=4$ selfdual theory are
\begin{subequations}
  \label{eq:neq4sdconstr}
\begin{align}
  \lbrace \mathcal{D}_{a \alpha'}, \mathcal{D}_{b \beta'}\rbrace &=   \epsilon_{\alpha' \beta'} W_{a b},\\
  [\mathcal{D}_{\alpha' a}, \mathcal{D}_{\beta \beta'}] &=   \epsilon_{\alpha' \beta'} \chi_{\beta a},\\
  [\mathcal{D}_{\alpha \alpha'}, \mathcal{D}_{\beta \beta'}] &= \epsilon_{\alpha' \beta'} F_{\alpha \beta}.
\end{align}
\end{subequations}  They can be written in an equivalent way as
\begin{gather}
  [\mathcal{D}_{\alpha}^{+}, \mathcal{D}_{\beta}^{+}] = 0,\qquad
  [\mathcal{D}_{a}^{+}, \mathcal{D}_{\alpha}^{+}] = 0,\qquad
  \lbrace \mathcal{D}_{a}^{+}, \mathcal{D}_{b}^{+}\rbrace = 0,\\
  [\mathcal{D}^{++}, \mathcal{D}_{a}^{+}] = 0,\qquad
  [\mathcal{D}^{++}, \mathcal{D}_{\alpha}^{+}] = 0,
\end{gather} where we have introduced a connection for the covariant derivative $D^{++}$ defined above.

We see that the derivatives $D_{\alpha}^{+}$, $D_{a}^{+}$, $D^{++}$ generate an integrable distribution.  Like explained in sec.~\ref{sec:general-philosophy}, we can solve the constraints involving the fermionic derivative $\mathcal{D}_{a}^{+}$ by going to a gauge where $A_{a}^{+} = 0$ and taking the remaining components $A_{\alpha}^{+}$ and $A^{++}$ to be such that $D_{a}^{+} A_{\alpha}^{+} = 0$ and $D_{a}^{+} A^{++} = 0$.  Such constraints are solved by restricting their dependence on superspace coordinates such that they depend on $(x^{\pm \alpha}, \theta^{+ a}, u^{\pm})$.

The remaining constraints can be written as a flatness condition.  If we define a $(0,1)$ connection $A = e^{- \alpha} A_{\alpha}^{+} + e^{--} A^{++}$ and introduce the Dolbeault operator $\bar{\partial} = e^{- \alpha} D_{\alpha}^{+} + e^{--} D^{++}$, then the constraints can be concisely written as $F \equiv \bar{\partial} A + A \wedge A = 0$.

In order to write an action which reproduces these constraints as equations of motion, we also define a $(3,0)$ form $\Omega = d^{4} \theta^{+}\; e^{+ \alpha} \wedge e_{\alpha}^{+} \wedge e^{++}$.  The action is then
\begin{equation}
  \label{eq:sd-action}
  S[A] = \int \Omega\wedge \tr \left(A \bar{\partial} A + \frac 2 3 A \wedge A \wedge A\right).
\end{equation}  The $U(1)$ charge cancels between the form $e^{+ \alpha} \wedge e_{\alpha}^{+} \wedge e^{++}$ and $d^{4} \theta^{+}$.

This is exactly the form of the action found by Witten in ref.~\cite{Witten:2003nn} as a twistor action.  In his notation $Z = (u^{+}, x^{+})$, $\bar{Z} = (u^{-}, x^{-})$ and $\psi = \theta^{+}$ and the connection $A$ depends on $(Z, \bar{Z}, \psi)$.  It is not hard to show that $\Omega \propto \epsilon_{i j k l} Z^{i} d Z^{j} d Z^{k} d Z^{l}$.  The variables $Z$ and $\bar{Z}$ are holomorphic and antiholomorphic coordinates on $\mathbb{CP}^{3}$, but in the $(u^{\pm},x^{\pm})$ parametrization only the $SL(2)_{L}$ symmetry is completely manifest.

In fact, this holomorphic Chern-Simons action has far more symmetry than just $SL(2)_{L}$.  Given that the action~\eqref{eq:sd-action} depends only on the complex structure, this means that any holomorphic change of coordinates is a symmetry.  We should note that if we write the action in components then the action of this symmetry group is obscured.  Also, one has to perform compensating gauge transformations in order to preserve the gauge.  In contrast, if we ask what transformations preserve the integrable distribution used to define the Dolbeault operator, the answer is easy. 

We should note that there are no local gauge invariant observables in the Chern-Simons formulation of this theory.  There are however holomorphic Wilson loops, which are discussed in more detail in sec.~\ref{sec:super-wilson-loops}.

How can we extract space-time fields from the connection $A$?  Let us define the following quantities
\begin{subequations}
  \label{eq:lin-twistor-transf}
\begin{align}
  \phi_{a b}(x, \theta) &= \int_{\mathbb{CP}^{1}} e^{++} \wedge D_{a}^{-} D_{b}^{-} A,\\
  \psi^{a \alpha'}(x, \theta) &= \frac 1 {3!} \epsilon^{a b c d} \int_{\mathbb{CP}^{1}} u^{+ \alpha'} e^{++} \wedge D_{b}^{-} D_{c}^{-} D_{d}^{-} A,\\
  G^{\alpha' \beta'}(x, \theta) &= \frac 1 {4!} \epsilon^{a b c d} \int_{\mathbb{CP}^{1}} u^{+ \alpha'} u^{+ \beta'} e^{++} \wedge D_{a}^{-} D_{b}^{-} D_{c}^{-} D_{d}^{-} A\label{eq:antisd}.
\end{align}
\end{subequations}  Using integration by parts, the algebra of covariant derivatives and the fact that $D_{a}^{+} A_{\alpha}^{+} = 0$ and $D_{a}^{+} A^{++} = 0$, it can be shown that these fields are invariant under Abelian gauge transformations\footnote{In the following we will restrict to gauge transformations $\lambda$ which do not depend explicitly on odd variables, but we keep the dependence on $u^{\pm}$ arbitrary.  It is natural to restrict to $D_{a}^{+} \lambda = 0$ in order to preserve the constraints $D_{a}^{+} A_{\alpha}^{+} = 0$ and $D_{a}^{+} A^{++} = 0$ but now we also require $D_{a}^{-} \lambda = 0$.  This restriction is not so great since gauge transformations of the superfields imply the gauge transformations of all the component fields.} $A \to A + \bar{\partial} \lambda$.  However, they are \emph{not} invariant under non-Abelian gauge transformations $A \to A + \bar{\partial} \lambda + [A, \lambda]$.  We should note that for fixed $x$ and $\theta$ the vielbein $e^{- \alpha}$ vanishes so we can replace $A$ by $e^{--} A^{++}$ in the equations above.  The vielbein $e^{--}$ can also be pulled through the covariant derivatives and brought next to $e^{++}$ where they form the measure on $\mathbb{CP}^{1}$.

Nevertheless, it is possible to add terms whose linearized gauge transformation cancels the nonlinear gauge transformation of the previous terms.  For example let us look for a term whose linearized gauge transformation cancels the gauge transformation of $\phi_{a b}$
\begin{equation}
  \delta \phi_{a b}(x, \theta) = \int_{\mathbb{CP}^{1}} e^{++} \wedge e^{--} [D_{a}^{-} D_{b}^{-} A^{++}, \lambda].
\end{equation}  In order to write the term which will cancel this gauge transformation, let us introduce the notation $(D^{++})^{-1}$ for the inverse of the operator $D^{++}$, when acting on functions of $u^{\pm}$.  Since $D^{++}$ has a kernel, $(D^{++})^{-1}$ is not unique.  We define $(D^{++})^{-1}$ when acting on a function $f^{++}(u)$ to be
\begin{equation}
  (D^{++})^{-1} f^{++}(u) = \int_{\mathbb{CP}^{1}} e^{++}(v) \wedge e^{--}(v)\; \frac {u^{+} v^{-}}{u^{+} v^{+}} f^{++}(v).
\end{equation}  It can be shown (see, for example, ref.~\cite[chap.~4]{Galperin:2001uw}) that $D^{++} (D^{++})^{-1} f^{++}(u) = f^{++}(u)$.  The difference $\lambda_{0} \equiv \lambda - (D^{++})^{-1} D^{++} \lambda$ is independent on $u$ (since it has charge zero and $D^{++} \lambda_{0} = 0$), but it may be nonzero.  This is not surprising since when taking the derivative $D^{++} \lambda$ we lose all the information about the zero mode of $\lambda$, i.e.\ the term of degree zero in the $u$ expansion.

Then, we have the linearized gauge transformation
\begin{multline}
  \delta_{\text{lin}} \left(-\int_{\mathbb{CP}^{1}} e^{++} \wedge e^{--} [D_{a}^{-} D_{b}^{-} A^{++}, (D^{++})^{-1} A^{++}]\right) =\\= -\int_{\mathbb{CP}^{1}} e^{++} \wedge e^{--}  [D_{a}^{-} D_{b}^{-} A^{++}, \lambda-\lambda_{0}].
\end{multline}

If we add these two terms, then the gauge transformation of $\phi_{a b}$ becomes
\begin{equation}
  \delta \phi_{a b} = [\phi_{a b}, \lambda_{0}] + \text{terms quadratic in $A^{++}$},
\end{equation} so $\lambda_{0}$ plays the role of space-time gauge transformation.  The terms quadratic in $A^{++}$ can be canceled by adding more correction terms.  We can now write the answer to all orders
\begin{equation}
\label{eq:nonabel}
  \phi_{a b}(x, \theta) = \int_{\mathbb{CP}^{1}} e^{++} \wedge e^{--} \sum_{p=1}^{\infty} [\underbrace{(D^{++})^{-1} A^{++}, \dotsc, [(D^{++})^{-1} A^{++}}_{p-1}, D_{a}^{-} D_{b}^{-} A^{++}] \cdots].
\end{equation}  The other space-time fields can be written similarly.

The other space-time superfields of interest are the bosonic and fermionic components of the superspace connection $A_{\alpha \alpha'}$ and $A_{\alpha' a}$.  In the Abelian theory they can be written as
\begin{subequations}
  \label{eq:gauge-lin-twistor-transf}
\begin{align}
  A_{\alpha \alpha'}(x, \theta) &= \int_{\mathbb{CP}^{1}} e^{++} \wedge   u_{\alpha'}^{-} D_{\alpha}^{-} A,\\
  A_{\alpha' a}(x, \theta) &= \int_{\mathbb{CP}^{1}} e^{++} \wedge u_{\alpha'}^{-} D_{a}^{-} A.
\end{align}
\end{subequations}  Here as well we can see that the only component of $A$ which contributes is the $A^{++}$ component (the $A_{\alpha}^{+}$ component is multiplied by the vielbein $e^{- \alpha}$ which vanishes for fixed $(x, \theta)$).  The new feature of these integrals with respect to the ones in eqs.~\eqref{eq:lin-twistor-transf} is the appearance of $u^{-}$ in the integrand.  This $u^{-}$ is responsible for the inhomogeneous term in the gauge transformations.

Under the linearized gauge transformation $\delta_{\text{lin}} A^{++} = D^{++} \lambda$ we can show after using the algebra of covariant derivatives and integration by parts that
\begin{equation}
\delta_{\text{lin}} A_{\alpha \alpha'} = \partial_{\alpha \alpha'} \lambda_{0}, \qquad
\delta_{\text{lin}} A_{\alpha' a} = 0,
\end{equation} where $\lambda_{0} = \int e^{++} \wedge e^{--} \lambda$.  The gauge parameter $\lambda_{0}$ is the same as the one found in the transformation of gauge covariant fields.  This can be shown as follows: an arbitrary function of the harmonic variables $u$ can be decomposed on an orthogonal basis of symmetrized products of $u^{\pm}$.  Integration over the $\mathbb{CP}^{1}$ projects on the zeroth order term in the expansion, while the rest of the terms vanish by orthogonality to the identity.

These space-time operators can be used to write the full (non-selfdual) $\mathcal{N}=4$ theory in twistor space.  The constraints for the full theory in space-time are
\begin{subequations}
  \label{eq:neq4constr}
\begin{align}
  \lbrace \mathcal{D}_{a \alpha'}, \mathcal{D}_{b \beta'}\rbrace &=   \epsilon_{\alpha' \beta'} W_{a b},\\
  [\mathcal{D}_{\alpha' a}, \mathcal{D}_{\beta \beta'}] &=   \epsilon_{\alpha' \beta'} \chi_{\beta a},\\
  [\mathcal{D}_{\alpha \alpha'}, \mathcal{D}_{\beta \beta'}] &= \epsilon_{\alpha' \beta'} F_{\alpha \beta} + \epsilon_{\alpha \beta} F_{\alpha' \beta'},
\end{align}
\end{subequations} where we have added an extra term $\epsilon_{\alpha \beta} F_{\alpha' \beta'}$ in the right-hand side of the commutator $[\mathcal{D}_{\alpha \alpha'}, \mathcal{D}_{\beta \beta'}]$.  Because of this extra term the constraints will not be writable as a flatness condition anymore.  Still, as we will see, the constraints can be written out explicitly in terms of the same two fields $A_{\alpha}^{+}$ and $A^{++}$ as before.

The constraints can be equivalently written in chiral harmonic twistor space as
\begin{equation}
  \label{eq:fulltheory}
  [\mathcal{D}^{+}_{a},\mathcal{D}^{+}_{b}]=0, \qquad
  [\mathcal{D}^{+}_{\alpha},\mathcal{D}^{+}_{a}]=0, \qquad
  [\mathcal{D}^{+}_{\alpha},\mathcal{D}^{+}_{\beta}]=\epsilon_{\alpha\beta}F^{++},
\end{equation}
where $F^{++}=u^{+\alpha'}u^{+\beta'}F_{\alpha'\beta'}$.  If we write the equations of motion in the Chalmers-Siegel form (see ref.~\cite{Chalmers:1996rq}), we need to set $F_{\alpha' \beta'} = g_{YM}^{2} G_{\alpha' \beta'}$, where $g_{YM}$ is the coupling constant and $G_{\alpha' \beta'}$ is an auxiliary field which, in the Abelian theory is written eq.~\eqref{eq:antisd}.  In the non-Abelian theory this expression is modified to make it gauge invariant as in eq.~\eqref{eq:nonabel}.

An action describing these equations of motion is necessarily not solely of the holomorphic Chern-Simons form, but has to be augmented by an additional term to include the local operator $F^{++}$.  This has the effect of adding an interaction term of the form $\ln \det (\bar{\partial} + A)$, as first suggested by Witten in ref.~\cite{Witten:2003nn}.  Adding all the terms we obtain the action of the full theory in chiral harmonic superspace
\begin{multline}
S_{F}[A] = \int d^{4} \theta^{+}\; \Omega\wedge \tr \left(A \bar{\partial} A + \frac 2 3 A \wedge A \wedge A\right) + \\ + g_{YM}^{2} \int d^{4}xd^{4}\theta^{+} d^{4} \theta^{-} \ln \det \left.(\bar{\partial} + A)\right\vert_{X},
\end{multline} where $X$ is the line in twistor space corresponding to the point $(x, \theta)$ in superspace.

\section{Super-Wilson loops}
\label{sec:super-wilson-loops}

In this section we review the construction of super-Wilson loops in twistor space (see ref.~\cite{Mason:2010yk} for the original paper and ref.~\cite{Adamo:2011pv} for a review).\footnote{In the abelian case Wilson loops in holomorphic Chern-Simons have been considered in refs.~\cite{1998math......6111T, 2001RSPTA.359.1413R, 2005math......2169F}.  The space-time version of the super-Wilson loop was constructed in ref.~\cite{CaronHuot:2010ek}.}  Later we will apply similar ideas to the $SU(3)/(U(1) \times U(1))$ coset formulation of $\mathcal{N}=4$ super-Yang-Mills theory.

In twistor space the harmonics parametrize a $\mathbb{CP}^{1}$ manifold and the full space on which the theory is formulated is $\mathbb{CP}^{3 \vert 4}$.  Space-time points correspond to lines, or $\mathbb{CP}^{1}$ embeddings in $\mathbb{CP}^{3 \vert 4}$.  We will generically denote such $\mathbb{CP}^{1} \subset \mathbb{CP}^{3 \vert 4}$ by $X$.

When restricted to a line $X$ the twistor connection $A$ is flat so there exists a gauge transformation $h$ satisfying
\begin{equation}
  \left.(\bar{\partial} + A(\sigma))\right\vert_{X} h(\sigma) = 0,
\end{equation} up to multiplication of $h$ by a constant group element to the right.  Then we define an analog of a Wilson line operator
\begin{equation}
  U(\sigma_{1}, \sigma_{0}) = h(\sigma_{1}) h(\sigma_{0})^{-1},
\end{equation} with properties
\begin{equation}
  U(\sigma, \sigma) = \mathbf{1}, \qquad
  U(\sigma_{1}, \sigma_{0})^{-1} = U(\sigma_{0}, \sigma_{1}), \qquad
  U(\sigma_{2}, \sigma_{0}) = U(\sigma_{2}, \sigma_{1}) U(\sigma_{1}, \sigma_{0}).
\end{equation}  Under a gauge transformation $(\bar{\partial} + A') = g (\bar{\partial} + A) g^{-1}$ the Wilson line operator transforms as
\begin{equation}
  U'(\sigma_{1}, \sigma_{0}) = g(\sigma_{1}) U(\sigma_{1}, \sigma_{0}) g(\sigma_{0})^{-1}.
\end{equation}

We can solve iteratively for $U$ in terms of the connection using
\begin{equation}
  U = \mathbf{1} + \bar{\partial}^{-1} (A U),
\end{equation} where $\bar{\partial}^{-1}$ acting on a $(0,1)$ form $f$ is defined by
\begin{equation}
  (\bar{\partial}^{-1} f)(\sigma) = \frac 1 \pi \int_{\mathbb{CP}^{1}} \left(\frac {d \sigma_{1}}{\sigma - \sigma_{1}} - \frac {d \sigma_{1}}{\sigma_{0} - \sigma_{1}}\right) \wedge f.
\end{equation}  This satisfies the boundary condition $(\bar{\partial}^{-1} f)(\sigma_{0}) = 0$.

Now we can explicitly write the expansion of $U$ as a power series in $A$:
\begin{multline}
  U(\sigma, \sigma_{0}) =
  \mathbf{1} + (\bar{\partial}^{-1} A)(\sigma) + \bar{\partial}^{-1} (A (\bar{\partial}^{-1} A))(\sigma) + \cdots =\\
  \mathbf{1} + \frac 1 {\pi} \int_{\mathbb{CP}^{1}} d \sigma_{1} \wedge A(\sigma_{1}) \frac {\sigma-\sigma_{0}}{(\sigma-\sigma_{1})(\sigma_{1}-\sigma_{0})} +\\
  \frac 1 {\pi^{2}} \int_{\mathbb{CP}^{1}} d \sigma_{1} \wedge A(\sigma_{1}) \int_{\mathbb{CP}^{1}} d \sigma_{2} \wedge A(\sigma_{2}) \frac {\sigma-\sigma_{0}}{(\sigma-\sigma_{2})(\sigma_{2}-\sigma_{1})(\sigma_{1}-\sigma_{0})} + \cdots.
\end{multline}

So far we defined a Wilson line operator between two points $\sigma_{0}$ and $\sigma$ on a line $X$.  The light-like Wilson loops of Mason and Skinner are defined as follows.  We have a contour $\mathcal{C}$ which is made up of pairwise intersecting lines $X_{i}$ such that two successive lines $X_{i}$ and $X_{i+1}$ intersect at a point in twistor space.  Each line $X_{i}$ has two distinguished points (whose local coordinates we denote by $\sigma_{i}$ and $\sigma_{i+1}$), where it intersects the previous line $X_{i-1}$ and the next line $X_{i+1}$.  Then, the supersymmetric Wilson loop of Mason and Skinner is defined as
\begin{equation}
  W = \tr (U_{X_{1}}(\sigma_{1}, \sigma_{2}) U_{X_{2}}(\sigma_{2}, \sigma_{3}) \cdots U_{X_{n}}(\sigma_{n}, \sigma_{1})).
\end{equation}

These Wilson loops are also useful when defining local operators.  For example, the local operator $\phi_{a b}(x, \theta)$ can be written as
\begin{equation}
  \label{eq:phi-wilson-line}
  \phi_{a b}(x, \theta) = \int_{X} e^{++}(\sigma) U_{X}(\tau, \sigma) (D_{a}^{-} D_{b}^{-} A)(\sigma) U_{X}(\sigma, \tau),
\end{equation} where $X$ is the line corresponding to $(x, \theta)$, $U_{X}$ is the Wilson line along $X$ and $(\sigma, \tau)$ are local coordinates on $X$.  Changing $\tau$ amounts to a global gauge transformation (see ref.~\cite{Adamo:2011dq} for a related discussion).

\section{Local operators}
\label{sec:local-operators}

We now turn to a discussion of (space-time) local gauge covariant operators in the $\mathcal{N}=3$ theory.  We will try to write down the scalar superfields $\phi_{i}(x, \theta, \bar{\theta})$ and $\bar{\phi}^{i}(x, \theta, \bar{\theta})$.

In order to write these space-time operators we need to eliminate the harmonic variables.  The way to do this is to integrate over them.  Recall that we denote the space of harmonics as $Q \subset \mathbb{CP}^{2} \times \mathbb{CP}^{2}$.  We normalize the integral over $Q$ by $\int_{Q} \text{vol} = 1$.

Just as in the twistor case we will first look for fields with zero charges under $U(1) \times U(1)$, with the right dimension and global symmetries and which are invariant under Abelian gauge transformations.

Before writing down the answers, we list some useful identities which can be proven by integration by parts
\begin{align}
  \int_{Q} u_{i}^{(1,0)} f^{(-1,0)} &= - \int_{Q} u_{i}^{(-1,1)} D^{(2,-1)} f^{(-1,0)} = - \int_{Q} u_{i}^{(0,-1)} D^{(1,1)} f^{(-1,0)},\\
  \int_{Q} u_{i}^{(0,-1)} f^{(0,1)} &= - \int_{Q} u_{i}^{(1,0)} D^{(-1,-1)} f^{(0,1)} = - \int_{Q} u_{i}^{(-1,1)} D^{(1,-2)} f^{(0,1)},\\
  \int_{Q} u_{i}^{(-1,1)} f^{(1,-1)} &= - \int_{Q} u_{i}^{(0,-1)} D^{(-1,2)} f^{(1,-1)} = - \int_{Q} u_{i}^{(1,0)} D^{(-2,1)} f^{(1,-1)},
\end{align} where $f$ is some arbitrary function.

There are several candidates for the scalar superfield $\phi_{i}$, with the right properties
\begin{subequations}
  \label{eq:lin-su3-transf}
\begin{align}
  &\int_{Q} \epsilon^{\alpha \beta} u_{i}^{(0,-1)} D_{\alpha}^{(-1,1)} D_{\beta}^{(-1,1)} A^{(2,-1)},\\
  &-\int_{Q} \epsilon^{\alpha \beta} u_{i}^{(-1,1)} D_{\alpha}^{(0,-1)} D_{\beta}^{(0,-1)} A^{(1,1)},\\
  &\int_{Q} \epsilon^{\alpha \beta} \left(-2 u_{i}^{(-1,1)} D_{\alpha}^{(0,-1)} D_{\beta}^{(-1,1)} A^{(2,-1)} + u_{i}^{(1,0)} D_{\alpha}^{(0,-1)} D_{\beta}^{(0,-1)} A^{(-1,2)}\right),\\
  &\int_{Q} \epsilon^{\alpha \beta} \left(2 u_{i}^{(0,-1)} D_{\alpha}^{(0,-1)} D_{\beta}^{(-1,1)} A^{(1,1)} - u_{i}^{(1,0)} D_{\alpha}^{(0,-1)} D_{\beta}^{(0,-1)} A^{(-1,2)}\right).
\end{align}
\end{subequations}  For the conjugate scalar superfield $\bar{\phi}^{i}$ we use the $\widetilde{\hspace{5mm}}$ conjugation and we also get four candidates.  Using integration by parts we can show that these fields are invariant under linearized gauge transformations $\delta A^{(p,q)} = D^{(p,q)} \lambda$.  In checking gauge invariance we can use the fact that the component fields $A^{(2,-1)}$, $A^{(1,1)}$ and $A^{(-1,2)}$ are analytic so when we apply $D_{\alpha}^{(1,0)}$ to them we obtain zero.

Note that these scalar superfield candidates are very similar to the expressions for the space-time fields in terms of twistor fields (see eq.~\eqref{eq:lin-twistor-transf}), at linearized level.

It is perhaps surprising that there are four candidates for the $\phi_{i}$ superfield.  However, when restricting on-shell and using the three linearized equations of motion we can show that all four candidates agree.  Off-shell, however, the four superfields in eq.~\eqref{eq:lin-su3-transf} are different.

In twistor space an axial gauge was used for quantizing the theory (see refs.~\cite{Mason:2010yk, Adamo:2011cb}).  The gauge condition in an axial gauge sets to zero a linear combination of the components of the twistor connection.  Since the three components of the gauge connection become dependent, in this gauge the cubic term $A^{3}$ in the Chern-Simons action vanishes.  As a result, the holomorphic Chern-Simons theory becomes free and the  Feynman rules simplify.

We can ask whether such a gauge is possible here.  It is not hard to see that this is not possible for a generic choice of the vector defining the axial gauge.  For example, if we set any of the components $A^{(2,-1)}$, $A^{(1,1)}$ or $A^{(-1,2)}$ to zero, we find that the scalar fields are set to zero, which is inconsistent.

A similar issue arises for the ambitwistor action of Mason and Skinner (see ref.~\cite{Mason:2005kn}), where the action is also of holomorphic Chern-Simons type, but formulated on ambitwistor space $\mathbb{A}_{[3]}$ which can be thought of as a quadric in $\mathbb{CP}^{3 \vert 3} \times \mathbb{CP}^{3 \vert 3}$.  In this case also it is not clear why the axial gauge can not be imposed.  Probably one way to understand why this gauge is inconsistent is by working out its counterpart in space-time as done above for the $\mathcal{N}=3$ formulation of $\mathcal{N}=4$ super-Yang-Mills.

\begin{figure}
  \centering
  \includegraphics{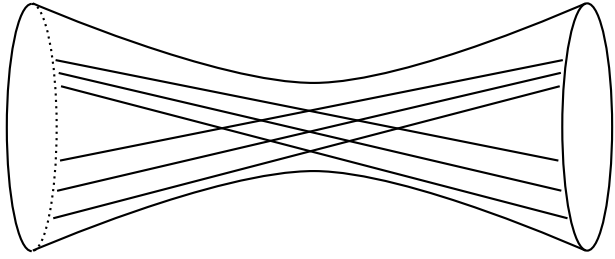}
  \caption{The coset $SU(3)/(U(1) \times U(1))$ can be represented as a quadric $Q \subset \mathbb{CP}^2 \times \mathbb{CP}^2$.  Two families of lines rule $Q$.  Each line in the first family sits at a fixed point in the second $\mathbb{CP}^{2}$ while each line in the second family sits at a fixed point in the first $\mathbb{CP}^{2}$.  Since the $\widetilde{\hspace{5mm}}$ conjugation swaps the two $\mathbb{CP}^{2}$, it also interchanges the two families of lines.  For lack of space we have not represented the $X_{3}$ submanifolds.}
  \label{fig:quadric}
\end{figure}

Let us now turn to writing down the local space-time operators which are invariant under nonlinear gauge transformations.  Clearly if we continue in the same spirit as for the selfdual theory, by adding correction terms, we will encounter great algebraic difficulties.  As we have seen, this is due to the fact that we have several candidates even for the linearized theory. Thus, as we go to higher orders, we have more and more possible correction terms to add.  This is a major difference with respect to the twistor formulation of $\mathcal{N}=4$ super-Yang-Mills, where fixing a point in superspace $(x, \theta)$ picks only one component of the twistor connection, namely $A^{++}$.  In that case it is natural to integrate over the line $X \subset \mathbb{CP}^{3 \vert 4}$ corresponding to $(x, \theta)$.

However, in the $\mathcal{N}=3$ formulation no such distinguished line exists, once we fix a point $(x, \theta, \bar{\theta})$ in superspace.  There are however three distinguished types of submanifolds.  The first one, denoted by $X_{1}$, is defined by setting $u_{i}^{1}$ and $u_{1}^{i}$ to be constant.  The submanifolds $X_{2}$ and $X_{3}$ are defined by analogy.  The submanifolds $X_{1}$ and $X_{2}$ are lines in the $\mathbb{CP}^{2} \times \mathbb{CP}^{2}$ which contains $Q = SU(3)/(U(1) \times U(1))$ (see fig.~\ref{fig:quadric}).

When imposing the defining constraints for $X_{1}$, $X_{2}$ and $X_{3}$ only two vielbeine survive, while the rest are set to zero by the constraints.  The nonvanishing ones are
\begin{alignat}{3}
  X_{1}: &\qquad e^{(-1,2)} &= u_{2}^{i} d u_{i}^{3}, &\qquad e^{(1,-2)} &= u_{3}^{i} d u_{i}^{2},\\
  X_{2}: &\qquad e^{(-2,1)} &= u_{1}^{i} d u_{i}^{3}, &\qquad e^{(2,-1)} &= u_{3}^{i} d u_{i}^{1},\\
  X_{3}: &\qquad e^{(-1,-1)} &= u_{1}^{i} d u_{i}^{2}, &\qquad e^{(1,1)} &= u_{2}^{i} d u_{i}^{1}.
\end{alignat}

The space of lines $X_{1}$ is parametrized by $\mathbb{CP}^{2}$, with coordinates $u_{i}^{1}$ and their complex conjugates $u_{1}^{i}$.  We denote this space by $Y_{1}$ and we define $Y_{2}$ and $Y_{3}$ by analogy.  These spaces come with natural volume forms,\footnote{In terms of homogeneous coordinates $[Z^{0}, \dotsc, Z^{n}]$ the volume form on $\mathbb{CP}^{n}$ can be written as \[\frac {\epsilon_{i_{0} \cdots i_{n}} Z^{i_{0}} d Z^{i_{1}} \wedge \cdots \wedge d Z^{i_{n}} \epsilon^{j_{0} \cdots j_{n}} \bar{Z}_{j_{0}} d \bar{Z}_{j_{1}} \wedge \cdots \wedge d \bar{Z}_{j_{n}}}{(Z \cdot \bar{Z})^{n+1}}.\]  Up to a constant multiplicative factor this is the same as $\Omega^{n}$, where $\Omega$ is the K\"ahler form $\Omega = \tfrac i 2 \partial \bar{\partial} \ln (Z \cdot \bar{Z})$.} which in terms of vielbeine can be written as
\begin{align}
  \mu_{Y_{1}} &= e^{(1,1)} \wedge e^{(2,-1)} \wedge e^{(-1,-1)} \wedge e^{(-2,1)}, \\
  \mu_{Y_{2}} &= e^{(1,-2)} \wedge e^{(-1,-1)} \wedge e^{(-1,2)} \wedge e^{(1,1)}, \\
  \mu_{Y_{3}} &= e^{(-2,1)} \wedge e^{(-1,2)} \wedge e^{(2,-1)} \wedge   e^{(1,-2)}.
\end{align}

Now we can write the local operators invariant under linearized gauge transformations as\footnote{The volume form on $X_{3}$ is $e^{(1,1)} \wedge e^{(-1,-1)}$ and we normalize the integrals $\int_{X_{3}} e^{(1,1)} \wedge e^{(-1,-1)} = 1$ and $\int_{Y_{I}} \mu_{Y_{I}} = 1$.  The differential forms appearing in the inner integral are pulled back from $Q$ to $X_{3}$.}
\begin{multline}
  \epsilon_{\alpha \beta} \phi_{i} = -\int_{Y_{3}} \mu_{Y_{3}} \int_{X_{3}} e^{(1,1)} \wedge u_{i}^{(-1,1)} D_{\alpha}^{(0,-1)} D_{\beta}^{(0,-1)} A =\\
  \int_{X_{1}} e^{(-1,2)} \wedge e^{(1,-2)} \int_{X_{2}} e^{(-2,1)} \wedge e^{(2,-1)} \int_{X_{3}} e^{(1,1)} \wedge u_{i}^{(-1,1)} D_{\alpha}^{(0,-1)} D_{\beta}^{(0,-1)} A.
\end{multline}  The result of the inner integration is invariant under linearized gauge transformations but it still depends, via $X_{3}$, on $u_{i}^{3}$ and $u_{3}^{i}$ variables.  The integral over $Y_{3}$ eliminates this dependence.

By the same reasoning as the one leading to eq.~\eqref{eq:phi-wilson-line}, the operator which is invariant under nonlinear gauge transformations\footnote{Here also we impose that these are proper i.e.\ not supergauge transformations.} is
\begin{multline}
  \label{eq:nonlin-twistor}
  \int_{X_{1}(\xi)} e^{(-1,2)}(\zeta) \wedge e^{(1,-2)}(\zeta) U_{X_{1}}(\xi, \zeta) \int_{X_{2}(\zeta)} e^{(-2,1)}(\tau) \wedge e^{(2,-1)}(\tau) U_{X_{2}}(\zeta, \tau) \\ \int_{X_{3}(\tau)} e^{(1,1)}(\sigma) \wedge U_{X_{3}}(\tau, \sigma) (u_{i}^{(-1,1)} D_{\alpha}^{(0,-1)} D_{\beta}^{(0,-1)} A)(\sigma) U_{X_{3}}(\sigma, \tau) U_{X_{2}}(\tau, \zeta) U_{X_{1}}(\zeta, \xi),
\end{multline} where $X_{1}(\xi)$ is the line $X_{1}$ containing the point\footnote{Here by abuse of notation we denote by $\xi$ the point in $Q$ and also its local coordinate on $X_{1}(\xi)$ and similarly for $\zeta$, $\tau$ and $\sigma$.} $\xi \in Q$, $X_{2}(\zeta)$ is the line $X_{2}$ containing the point $\zeta \in Q$ and $X_{3}(\tau)$ is the line $X_{3}$ containing the point $\tau \in Q$.  The Wilson line $U_{X_{I}}$ can be constructed as in sec.~\ref{sec:super-wilson-loops} in terms of the connection $A$ and the inverse of the Dolbeault operator restricted to $X_{I}$.  The definition in eq.~\eqref{eq:nonlin-twistor} depends on a choice of a reference point $\xi \in Q$.

Just like in the linearized case, there are several ways (in fact an infinite number!) to construct local space-time fields starting from the connection $A$ on harmonic superspace.  On-shell the connection $A$ is flat on $Q$ so the contours for the Wilson lines can be freely deformed.  This implies that all the different representations of the local space-time operators in terms of the connection $A$ are equivalent on-shell.

\section{Conclusions}
\label{sec:conclusions}

The formalism developed here allows us in principle to compute physical quantities while preserving a large amount of off-shell supersymmetry. One natural class of observables are correlation functions of gauge-invariant local operators. In harmonic language such correlation functions become Wilson loops with operator insertions as we showed in sec.~\ref{sec:local-operators}. It may be interesting to see whether it would be easier to compute anomalous dimensions in this formalism and how integrability manifests itself. One thing to understand here would be how dualities between scattering  amplitudes, Wilson loops and correlation functions are realized in this harmonic language.
\medskip

However, we are left with numerous questions.

First of all, we would like to understand how to quantize the $\mathcal{N}=3$ theory.  We have mentioned above the similarity with the ambitwistor action which makes us confident that the difficulties in quantization are the same in both cases. In general, the study of non-chiral harmonic superspace may shed some light on the more mysterious ambitwistor formulation of $\mathcal{N}=4$ SYM.

In ref.~\cite{Delduc:1988cp} the two-point function of the gauge fields was computed.  However, in the gauge of ref.~\cite{Delduc:1988cp} the ghosts do not decouple, which complicates the Feynman rules.  Is it possible to impose an algebraic gauge condition where the ghosts decouple?

Another vexing problem is the problem of regularization.  These Chern-Simons-type actions are finite but some interesting ``observables'' require regularization.  So far a lot could be computed while mostly ignoring regularization issues.  Nevertheless, we should strive to obtain these results rigorously.  One regularization proposal has been put forward in ref.~\cite{Heckman:2011ju}, but so far it has not been used for explicit computations.

What other terms can be added to the action which are gauge
invariant and also invariant under $SU(2,2\vert 3)$?  This is related to the question, to our knowledge still unsolved, of how to write the analog of the $\theta$ angle in the twistor case.

For usual Chern-Simons one can make gauge transformations by a group
element $g$ which is not continuously connected to the identity, in which case the action transforms by an additive term.  It is unclear to us what happens in the holomorphic Chern-Simons case.

All of the constructions presented in this paper require an integrable distribution of rank $(3 \vert \kappa)$, where $\kappa$ is the odd rank of the distribution.  We can naturally ask what would be the space-time interpretation of higher holomorphic Chern-Simons forms, which can be constructed in all odd dimensions.

As explained in ref.~\cite{Dijkgraaf:1989pz}, there are other subtleties in the construction of Chern-Simons theory if the gauge group is not connected or simply connected.  It would be interesting to see if such subtleties also occur for holomorphic Chern-Simons theories or the ones built from a CR structure.

We should mention here a curious formulation of $\mathcal{N}=4$ five-dimensional super-Yang-Mills obtained by Sokatchev in ref.~\cite{Sokatchev:1988qr}.  The off-shell formulation in ref.~\cite{Sokatchev:1988qr} is also of Chern-Simons type which hints that five-dimensional super-Yang-Mills might be finite~\cite{Douglas:2010iu,Lambert:2010iw,Kallen:2012zn,Bern:2012di}.  In order to investigate this more closely one would at least want to study the compactification of this theory on a circle.  Another useful test would be to compute the partition function on $\mathbf{S}^{5}$.  However, the off-shell formulation is only available for the theory defined on flat space-time $\mathbb{R}^{1,4}$.

Another interesting question regards theories with fewer supersymmetries.  Mason and Skinner gave an ambitwistor formulation for pure Yang-Mills, whose ambitwistor Lagrangian contains a triple derivative of a delta function $\delta'''(Z \cdot W)$.  The fact that the equations of motion of pure Yang-Mills theory correspond to extensions to the triple neighborhood about the $Z \cdot W = 0$ locus has been known before from work by Witten~\cite{Witten:1978xx} and Isenberg, Iasskin and Green~\cite{Isenberg:1978kk}.  The derivatives of the delta functions serve to cancel the twistor scaling which in the supersymmetric case was canceled by integrations over odd variables.  We expect that a similar mechanism is at work in the construction based on the $SU(3)/(U(1) \times U(1))$ coset.

In ref~\cite{Berkovits:2004tx} a Chern-Simons type string field theory
was proposed which describes scattering amplitudes of $\mathcal{N}=4$
SYM coupled to $\mathcal{N}=4$ conformal supergravity. There are two types of
interaction terms. One is the usual cubic term of Chern-Simons
theories while the other contains an insertion of a spectral flow operator. At loop
level we encounter the usual difficulties with conformal supergravity,
so this action has not been used for any explicit loop-level
computations. Nevertheless, this is an exotic example of solving the
constraints in a way which is different from the usual harmonic
approach.

Finally, another important question would be to understand the analogs of the constructions we presented for gravity.  The analog of holomorphic Chern-Simons for gravity was discussed in ref.~\cite{Mason:2007ct}, where the gauge group was identified with the group of holomorphic Poisson transformations of supertwistor space, but this only describes the selfdual supergravities.  An understanding of non-selfdual theories may be possible by reexamining the lessons of the harmonic approach.

\section{Acknowledgments}

CV would like to thank David Skinner for discussions about ambitwistors in Santa Barbara during the KITP program ``The Harmony of Scattering Amplitudes''.  We are grateful to Niklas Beisert and Matteo Rosso for discussions on related subjects. The work of BUWS is supported in part by grant no.~200021-137616 from the Swiss National Science Foundation.

\appendix

\section{Conventions}
\label{sec:conventions}

We raise and lower indices with the antisymmetric tensor $\epsilon$, $u_{\alpha} = \epsilon_{\alpha \beta} u^{\beta}$, $u^{\alpha} = \epsilon^{\alpha \beta} u_{\beta}$.  Raising and then lowering an index of a spinor leaves the spinor invariant, so we have $\epsilon_{\alpha \beta} e^{\beta \gamma} = \delta_{\alpha}^{\gamma}$ and $\epsilon^{\alpha \beta} \epsilon_{\beta \gamma} = \delta_{\gamma}^{\alpha}$.

Whenever we need explicit forms for the $\epsilon$ tensors we use
\begin{equation}
  \epsilon_{\cdot \cdot} =
  \begin{pmatrix}
    0 & 1\\-1 & 0
  \end{pmatrix}, \qquad
  \epsilon^{\cdot \cdot} =
  \begin{pmatrix}
    0 & -1\\1 & 0
  \end{pmatrix}.
\end{equation}

\section{Coset space generalities}
\label{sec:coset-spaces}

Let $G$ be a Lie group and $H$ a subgroup and let $\mathfrak{g}$ and $\mathfrak{h}$ be the corresponding Lie algebras.  We denote the generators of $\mathfrak{h}$ by $X_{i}$ and the remaining generators in $\mathfrak{g}$ by $Y_{\alpha}$.

We parametrize the coset $G/H$ by $\Omega = \exp (\xi^{\alpha} Y_{\alpha}) \in G$ and $\xi$ can be seen as coordinates on the coset manifold.  The action of $G$ on the coset is given by
\begin{equation}
  g \exp(\xi^{\alpha} Y_{\alpha}) = \exp(\xi'^{\alpha}(\xi, g) Y_{\alpha}) h(\xi, g),
\end{equation} where $g \in G$ and $h(\xi, g) \in H$.

For a matrix group we can form the quantity $\Omega^{-1} d \Omega \in \mathfrak{g}$, which can then be decomposed as
\begin{equation}
  \Omega^{-1} d \Omega = e^{\alpha} Y_{\alpha} + \omega^{i} X_{i}.
\end{equation}  Under a transformation by $g$, we have $\Omega \to \Omega' = g \Omega h^{-1}$ and for the components
\begin{align}
  e'^{\alpha} Y_{\alpha} &= h e^{\alpha} Y_{\alpha} h^{-1},\\
  \omega'^{i} X_{i} &= h \omega^{i} X_{i} h^{-1} + h d h^{-1}.
\end{align}  The transformation properties justify our identification of $e$ as vielbeine and of $\omega$ as connections.

We can compute the derivatives of $e$ and $\omega$ and we get
\begin{multline}
  d e^{\alpha} Y_{\alpha} + d \omega^{i} X_{i} = d (\Omega^{-1} d \Omega) = -\frac 1 2 e^{\alpha} \wedge e^{\beta} [Y_{\alpha}, Y_{\beta}] -\\ e^{\alpha} \wedge \omega^{i} [Y_{\alpha}, X_{j}] - \frac 1 2 \omega^{i} \wedge \omega^{j} [X_{i}, X_{j}].
\end{multline}

Define the structure constants of the algebra $\mathfrak{g}$ by
\begin{align}
  [X_{i}, X_{j}] &= f_{i j}^{k} X_{k}, \\
  [X_{i}, Y_{\alpha}] &= f_{i \alpha}^{\beta} Y_{\beta}, \\
  [Y_{\alpha}, Y_{\beta}] &= f_{\alpha \beta}^{i} X_{i} + f_{\alpha \beta}^{\gamma} Y_{\gamma}.
\end{align}  Using this we can write down the derivatives of $e^{\alpha}$ and $\omega^{i}$ very explicitly as
\begin{align}
  d e^{\alpha} + e^{\beta} \wedge \omega^{i} f_{\beta i}^{\alpha} &=   -\frac 1 2 f_{\beta \gamma}^{\alpha} e^{\beta} \wedge e^{\gamma},\\
  d \omega^{i} + \frac 1 2 \omega^{j} \wedge \omega^{k} f_{j k}^{i} &= -\frac 1 2 e^{\alpha} \wedge e^{\beta} f_{\alpha \beta}^{i}.
\end{align}  The first equation above gives the covariant derivative of $e$.  When $f_{\beta \gamma}^{\alpha} \neq 0$, the connection we defined has torsion.  Only when $[Y, Y] \sim X$ the connection does not have torsion.  The second equation gives the curvature.

We should note that, since $[X, Y] \sim Y$, whenever we make a transformation by $g \in H$, we have that $h(\xi, g) = g$.

Let us now introduce a class of functions (or fields) on $G/H$ with the following transformation properties under $G$
\begin{equation}
  \phi'(\xi'(\xi, g)) = \rho(h(\xi, g)) \cdot \phi(\xi),
\end{equation} where $\xi'$ and $h$ have been defined above and $\rho$ is a representation of $H$ and $\phi(\xi)$ transforms under this representation.  For example, the quantity $e^{\alpha} Y_{\alpha}$ transforms in this way under the adjoint representation.  It is then easy to see that the covariant derivative defined as
\begin{equation}
  D \phi = (d + \omega^{i} \rho(X_{i})) \cdot \phi
\end{equation} transforms in the same way as $\phi$.  This covariant derivative can be decomposed on the vielbeine as $D = e^{\alpha} D_{\alpha}$, this decomposition defining the components $D_{\alpha}$ of the covariant derivative.

Using the vielbeine we can also construct an integration measure $\lvert \mu \rvert$ on the coset by taking
\begin{equation}
  \lvert \mu \rvert = d^{n} \xi \det e,
\end{equation} where the coset is $n$-dimensional and $\xi$ are the coordinates parametrizing it and $e$ is the vielbein matrix $e_{\beta}^{\alpha}$ extracted from $e^{\alpha} = e_{\beta}^{\alpha} d \xi^{\beta}$.  We can also write the measure as a top form
\begin{equation}
  \mu = e^{1} \wedge \dotso \wedge e^{n}.
\end{equation}

Let us discuss the example of $SU(2)/U(1)$ coset.  We set
\begin{equation}
  \Omega = \begin{pmatrix} u^{+1} & u^{-1}\\ u^{+2} & u^{-2}\end{pmatrix},
\end{equation} with $u^{+1} u^{-2} - u^{+2} u^{-1} = 1$ and $u^{-1} = -(u^{+2})^{*}$ and $u^{-2} = (u^{+1})^{*}$.  The $U(1)$ action is $u^{\pm} \to e^{\pm i \phi} u^{\pm}$.  It can be embedded in $SU(2)$ as $\left(\begin{smallmatrix} e^{i \phi} & 0\\0 & e^{-i \phi}\end{smallmatrix}\right)$ and it acts to the right on $\Omega$.

Then, computing $\Omega^{-1} d \Omega$ we find the vielbeine and connection
\begin{align}
  e^{--} &= -u^{-1} d u^{-2} + u^{-2} d u^{-1} = u_{\alpha'}^{-} d u^{- \alpha'},\\
  e^{++} &= u^{+1} d u^{+2} - u^{+2} d u^{+1} = -u_{\alpha'}^{+} d u^{+ \alpha'},\\
  \omega^{0} &= -u^{-1} d u^{+2} + u^{-2} d u^{+1} = -u^{+1} d u^{-2} + u^{+2} d u^{-1} = -u_{\alpha'}^{-} d u^{+ \alpha'} = - u_{\alpha'}^{+} d u^{- \alpha'},
\end{align} where we have indicated the charges of the vielbeine under the $U(1)$ group.

Since the field $H$ is $U(1)$ in this case, the representations are multiplication by phases $e^{i q \phi}$.  Then, the covariant derivative when acting on a function $f^{(q)}$ of charge $q$, reads
\begin{equation}
  D f^{(q)} = \left(d u^{+ \alpha'} \frac {\partial}{\partial u^{+ \alpha'}} + d u^{- \alpha'} \frac {\partial}{\partial u^{- \alpha'}} + q \omega^{0}\right) f^{(q)}.
\end{equation}  Given the transformation of $u^{\pm}$ under the $U(1)$ charge, we have that the the function $f^{(q)}$ is homogeneous so
\begin{equation}
  \left(u^{+ \alpha'} \frac {\partial}{\partial u^{+ \alpha'}} - u^{- \alpha'} \frac {\partial}{\partial u^{- \alpha'}}\right) f^{(q)} = q f^{(q)}.
\end{equation}  Using this in the expression of the covariant derivative we find
\begin{equation}
  D = e^{++} D^{--} + e^{--} D^{++},
\end{equation} with
\begin{equation}
  D^{++} = u^{+ \alpha'} \frac {\partial}{\partial u^{- \alpha'}}, \qquad
  D^{--} = u^{- \alpha'} \frac {\partial}{\partial u^{+ \alpha'}}.
\end{equation}  These covariant derivatives are dual to the vielbeine we constructed
\begin{equation}
  \langle D^{\pm \pm}, e^{\mp \mp}\rangle = 1, \qquad
  \langle D^{\pm \pm}, e^{\pm \pm}\rangle = 0.
\end{equation}

The top form on the coset $SU(2)/U(1)$ is given by
\begin{equation}
  \mu = u_{\alpha'}^{+} d u^{+ \alpha'} \wedge u_{\beta'}^{-} d u^{- \beta'}.
\end{equation}

Let us now discuss the $SU(3)/(U(1) \times U(1))$ coset.  Following Galperin at al.\ we parametrize the $SU(3)$ group by a $3 \times 3$ matrix $u_{i}^{I}$, with $i = 1,2,3$ and while $I$ is labeled by the $U(1) \times U(1)$ charges $I = (1,0), (0,-1), (-1,1)$.  Occasionally it will be convenient to use a short notation and then we will let $I$ range over $1,2,3$, with the understanding that these labels correspond to the charges $(1,0), (0,-1), (-1,1)$.  Under the $U(1) \times U(1)$ transformations these matrix elements $u$ transform as
\begin{alignat}{2}
  u_{1}^{i} &\to e^{-i \phi_{1}} u_{1}^{i}, &\qquad
  u_{i}^{1} &\to e^{i \phi_{1}} u_{i}^{1},\\
  u_{2}^{i} &\to e^{i \phi_{2}} u_{2}^{i}, &\qquad
  u_{i}^{2} &\to e^{-i \phi_{2}} u_{i}^{2},\\
  u_{3}^{i} &\to e^{i (\phi_{1}-\phi_{2})} u_{3}^{i}, &\qquad
  u_{i}^{3} &\to e^{-i (\phi_{1}-\phi_{2})} u_{i}^{3}.
\end{alignat}

The inverse matrix is denoted by $u_{I}^{i}$ and therefore we have $u_{i}^{I} u_{I}^{j} = \delta_{i}^{j}$ and $u_{I}^{i} u_{i}^{J} = \delta_{I}^{J}$.  Since the matrix $u$ is unitary we also have $(u_{i}^{I})^{*} = u_{I}^{i}$.  Finally, we have $\det u_{i}^{I} = 1$.  Since $u_{i}^{I}$ has unit determinant we can write the inverse explicitly $u_{1}^{i} = \epsilon^{i j k} u_{j}^{2} u_{k}^{3}$ and cyclicly related identities.  Then, if we take the differential of $\epsilon^{i j k} u_{i}^{1} u_{j}^{2} u_{k}^{3} = 1$ and we use the formula for the inverse, we obtain $u_{I}^{i} d u_{i}^{I} = 0$.

Now, we compute the one-form $u_{J}^{i} d u_{i}^{J}$ and we find the vielbeine
\begin{alignat}{2}
  e^{(-1,-1)} &= u_{1}^{i} d u_{i}^{2}, &\qquad
  e^{(1,1)} &= u_{2}^{i} d u_{i}^{1},\\
  e^{(-2,1)} &= u_{1}^{i} d u_{i}^{3}, &\qquad
  e^{(2,-1)} &= u_{3}^{i} d u_{i}^{1},\\
  e^{(-1,2)} &= u_{2}^{i} d u_{i}^{3}, &\qquad
  e^{(1,-2)} &= u_{3}^{i} d u_{i}^{2}.
\end{alignat}  The connections $\omega$ can be chosen to be
\begin{align}
  \omega_{1} &= -u_{1}^{i} d u_{i}^{1} = u_{2}^{i} d u_{i}^{2} + u_{3}^{i} d u_{i}^{3},\\
  \omega_{2} &= u_{2}^{i} d u_{i}^{2} = -u_{1}^{i} d u_{i}^{1} - u_{3}^{i} d u_{i}^{3}.
\end{align}  These choices for $\omega$ have the advantage that their transformations under $U(1) \times U(1)$ are simple
\begin{align}
  \omega_{1} &\to \omega_{1} - i d \phi_{1},\\
  \omega_{2} &\to \omega_{2} - i d \phi_{2}.
\end{align}

Let us compute the covariant derivative when acting on functions with charges $(q_{1}, q_{2})$ under $U(1) \times U(1)$.  According to the general theory presented above, we have
\begin{equation}
  D f^{(q_{1}, q_{2})} = \left(d u_{i}^{I} \frac {\partial}{\partial u_{i}^{I}} + q_{1} \omega_{1} + q_{2} \omega_{2}\right) f^{(q_{1}, q_{2})}.
\end{equation}  The homogeneity properties imply
\begin{align}
  \left(u_{i}^{1} \frac {\partial}{\partial u_{i}^{1}} - u_{i}^{3} \frac {\partial}{\partial u_{i}^{3}}\right) f^{(q_{1}, q_{2})} &= q_{1} f^{(q_{1}, q_{2})},\\
  \left(-u_{i}^{2} \frac {\partial}{\partial u_{i}^{2}} + u_{i}^{3} \frac {\partial}{\partial u_{i}^{3}}\right) f^{(q_{1}, q_{2})} &= q_{2} f^{(q_{1}, q_{2})}.
\end{align}  Then, after using the following equalities
\begin{align}
  d u_{i}^{1} &= u_{i}^{3} e^{(2,-1)} + u_{i}^{2} e^{(1,1)} -   u_{i}^{1} \omega_{1},\\
  d u_{i}^{2} &= u_{i}^{1} e^{(-1,-1)} + u_{i}^{3} e^{(1,-2)} +   u_{i}^{2} \omega_{2},\\
  d u_{i}^{3} &= u_{i}^{1} e^{(-2,1)} + u_{i}^{2} e^{(-1,2)} + u_{i}^{3} (\omega_{1} - \omega_{2}),
\end{align} in the formula for the covariant derivative, we find
\begin{multline}
  D = \sum_{(q_{1}, q_{2})} e^{(q_{1}, q_{2})} D^{(-q_{1}, -q_{2})} = e^{(2,-1)} u_{i}^{3} \frac \partial {\partial u_{i}^{1}} + e^{(1,1)} u_{i}^{2} \frac \partial {\partial u_{i}^{1}} + e^{(1,-2)} u_{i}^{3} \frac \partial {\partial u_{i}^{2}} +\\+ e^{(-1,2)} u_{i}^{2} \frac \partial {\partial u_{i}^{3}} + e^{(-1,-1)} u_{i}^{1} \frac \partial {\partial u_{i}^{2}} + e^{(-2,1)} u_{i}^{1} \frac \partial {\partial u_{i}^{3}}.
\end{multline}  These covariant derivatives are dual to the vielbeine
\begin{equation}
  \langle D^{(-q_{1}, -q_{2})}, e^{(r_{1}, r_{2})}\rangle = \delta_{q_{1}, r_{1}} \delta_{q_{2} r_{2}}.
\end{equation}

Under complex conjugation $(u_{i}^{I})^{*} = u_{I}^{i}$ we have
\begin{gather}
  (\omega_{1})^{*} = -\omega_{1}, \qquad
  (\omega_{2})^{*} = -\omega_{2}, \\
  (e^{(q_{1}, q_{2})})^{*} = -e^{(-q_{1}, -q_{2})}, \qquad
  (D^{(q_{1}, q_{2})})^{*} = -D^{(-q_{1}, -q_{2})}.
\end{gather}

We now list the derivatives of the vielbeine
\begin{align}
  D e^{(-1,-1)} &\equiv d e^{(-1,-1)} + (-\omega_{1} - \omega_{2}) \wedge e^{(-1,-1)} = -e^{(-2,1)} \wedge e^{(1,-2)},\\
  D e^{(-2,1)} &\equiv d e^{(-2,1)} + (-2 \omega_{1} + \omega_{2}) \wedge e^{(-2,1)} = -e^{(-1,-1)} \wedge e^{(-1,2)},\\
  D e^{(-1,2)} &\equiv d e^{(-1,2)} + (-\omega_{1} + 2 \omega_{2})   \wedge e^{(-1,2)} = -e^{(1,1)} \wedge e^{(-2,1)},\\
  D e^{(1,1)} &\equiv d e^{(1,1)} + (\omega_{1} + \omega_{2}) \wedge   e^{(1,1)} = - e^{(-1,2)} \wedge e^{(2,-1)},\\
  D e^{(2,-1)} &\equiv d e^{(2,-1)} + (2 \omega_{1} - \omega_{2})   \wedge e^{(2,-1)} = -e^{(1,-2)} \wedge e^{(1,1)},\\
  D e^{(1,-2)} &\equiv d e^{(1,-2)} + (\omega_{1} - 2 \omega_{2}) \wedge e^{(1,-2)} = -e^{(2,-1)} \wedge e^{(-1,-1)}.
\end{align}  The covariant derivatives satisfy commutation relations which are dual to these relations
\begin{subequations}
  \label{eq:su3-cov-algebra}
\begin{alignat}{2}
  [D^{(-2,1)}, D^{(1,-2)}] &= D^{(-1,-1)}, &\quad
  [D^{(-1,-1)}, D^{(-1,2)}] &= D^{(-2,1)},\\
  [D^{(1,1)}, D^{(-2,1)}] &= D^{(-1,2)}, &\quad
  [D^{(-1,2)}, D^{(2,-1)}] &= D^{(1,1)},\\
  [D^{(1,-2)}, D^{(1,1)}] &= D^{(2,-1)}, &\quad
  [D^{(2,-1)}, D^{(-1,-1)}] &= D^{(1,-2)}.
\end{alignat}
\end{subequations}

\begin{figure}
  \centering
  \includegraphics{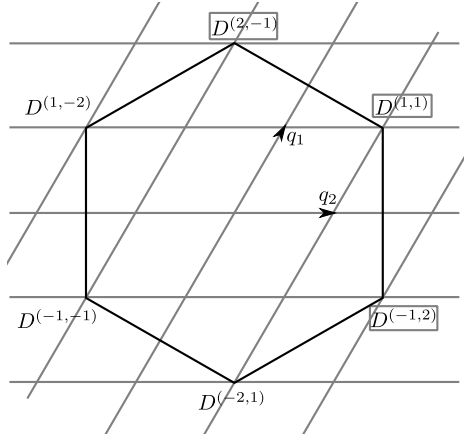}
  \caption{The covariant derivatives of the coset $SU(3)/(U(1) \times U(1))$.  We can choose an integrable distribution in several ways.  One choice is marked by boxes, but other possibilities can be obtained by rotation.}
  \label{fig:su3}
\end{figure}

\section{CR manifolds}
\label{sec:cr-manifolds}

The cosets we are using have a CR structure which is essential in the construction of the action and in writing the equations of motion.  We will review the essential points of the construction below, illustrating the definitions by examples of interest.  A good reference for the material in this section is the book~\cite{Boggess:1991}.  A short but useful discussion can also be found in refs.~\cite{Howe:1995md, Mason:2005kn}.

We say that a manifold $M$ with dimension $2 n + m$ has a CR structure if it has a rank $n$ distribution $L$ of the complexified tangent bundle of $M$ and $L \cap \bar{L} = 0$.

The coset $SU(3)/(U(1) \times U(1))$ has a CR structure.  The integrable distribution $L$ is generated by $D^{(-2,1)}$, $D^{(1,-2)}$, $D^{(-1,-1)}$.  The integrability follows from
\begin{equation}
  [D^{(-2,1)}, D^{(1,-2)}] = D^{(-1,-1)}, \quad
  [D^{(-2,1)}, D^{(-1,-1)}] = 0, \quad
  [D^{(1,-2)}, D^{(-1,-1)}] = 0.
\end{equation} The distribution $\bar{L}$ is generated by $D^{(2,-1)}$, $D^{(1,1)}$, $D^{(-1,2)}$.  Since these vector fields are all independent, it follows that $L \cap \bar{L} = 0$.

Starting with a CR structure we can construct a Dolbeault operator as follows.  By definition, $L$ is a subbundle of the complexified tangent bundle of $M$.  The subbundle $L$ has a dual bundle $L^{*}$ of one-forms and moreover there is projection $\pi$ from the cotangent bundle of $M$ to $L^{*}$.  Using these ingredients we can define the Dolbeault operator $\bar{\partial}$ acting on a function $f$ by
\begin{equation}
  \bar{\partial} f = \pi d f.
\end{equation}  Then the action of $\bar{\partial}$ on $(0,q)$-forms can be defined such that the usual rules of differential calculus apply.  The definition is the same as above, but now $\pi$ projects to $\bigwedge^{(q+1)} L^{*}$.

In the $SU(3)/(U(1) \times U(1))$ case with $L$ generated by $D^{(-2,1)}$, $D^{(1,-2)}$, $D^{(-1,-1)}$, we have that $L^{*}$ is generated by $e^{(2,-1)}$, $e^{(-1,2)}$ and $e^{(1,1)}$.  The projection $\pi$ sends the one-forms $\omega_{1}$, $\omega_{2}$, $e^{(-2,1)}$, $e^{(1,-2)}$ and $e^{(-1,-1)}$ to zero.  Therefore, when acting on a function $f$ we have
\begin{equation}
  \bar{\partial} f = \left(e^{(-2,1)} D^{(2,-1)} + e^{(1,-2)} D^{(-1,2)} + e^{(-1,-1)} D^{(1,1)}\right) f.
\end{equation}

The action on the one-forms is very simple.  We only list here the action on the one-forms $e^{(q_{1}, q_{2})}$
\begin{alignat}{2}
  \bar{\partial} e^{(-1,-1)} &= -e^{(-2,1)} \wedge e^{(1,-2)}, &\qquad
  \bar{\partial} e^{(-2,1)} &= 0,\\
  \bar{\partial} e^{(-1,2)} &= 0, &\qquad
  \bar{\partial} e^{(1,1)} &= 0,\\
  \bar{\partial} e^{(2,-1)} &= 0, &\qquad
  \bar{\partial} e^{(1,-2)} &= 0.
\end{alignat}

Using the integrability of $L$ it can be shown that $\bar{\partial}^{2} = 0$.  In the case of the coset $SU(3)/(U(1) \times U(1))$ this can also be checked explicitly.

\section{Killing vectors for \texorpdfstring{$\mathcal{N}=3$}{N=3} harmonic superspace}
\label{sec:kill-vect}

Killing vectors for $\mathcal{N}$-extended harmonic superspaces with a CR structure can be effectively calculated using the algorithm given in \cite{Howe:1995md}. Let $V=V_{0} + V_{u}$ be a Killing vector with
\[V_{0} = F^{\alpha\dot\alpha}\del_{\alpha\dot\alpha} + f^{i\alpha}D_{i\alpha} - \bar f^{\dot\alpha}_{i}\bar{D}^{i}_{\dot\alpha},\qquad V_{u} =  f^{(-q_{1},-q_{2})}D^{(q_{1},q_{2})}\] where $D^{(q_{1},q_{2})}$ are the harmonic derivatives. Following \cite{Howe:1995md} the functions $F$ and $\bar f$ can be shown to satisfy
\begin{align}\del_{(\alpha}^{(\dot\alpha}F_{\beta)}^{\dot\beta)}=0,\qquad D_{i\alpha}F_{\beta\dot\beta} = - i \epsilon_{\alpha\beta}\bar{f}_{i\dot\beta}\label{eq:killing}\end{align}
and 
\[f^{(-q_{1},-q_{2})} = f^I_{J}=\frac12 (D_{J\alpha})f^{\alpha I}-\frac13\delta^{I}_{J}D_{\alpha K}f^{\alpha K}\] in the notation given above. 

The components of the Killing vector are constrained by the requirement that a superconformal transformation preserves the CR structure under commutation. Given the distribution of $\mathcal{N}=3$ harmonic superspace \[\{D^{(1,0)}_{\alpha},\bar D^{(0,1)}_{\dot\alpha},D^{(2,-1)},D^{(-1,2)},D^{(1,1)}\}\] the conditions on $V$ are
\begin{align}
  \label{eq:Killing}
  [D^{(p,q)},V]&=0,\\
  [D^{(1,0)}_{\alpha},V]&\propto D^{(1,0)}_{\alpha},D^{(2,-1)},D^{(1,1)}\\
  [\bar D^{(0,1)}_{\dot\alpha},V]&\propto \bar D^{(0.1)}_{\dot\alpha},D^{(-1,2)},D^{(1,1)}
\end{align}
The first condition implies that the Killing vector is uncharged under $U(1)\times U(1)$ and the components of $V_{0}$ are independent of the harmonics. It further implies some relations among the components of $V_{u}$, i.e. \[D^{(p,q)}f^{(-k,-l)} \pm f^{(p-k,q-l)}=0\] whenever the superscript $(p-k,q-l)$ is an allowed combination of weights, otherwise the second term is zero. Finally, the last two constraints imply analyticity of some of the components of the vector fields $V_{0}$ and $V_{u}$.

These component functions have mass dimensions 
\[[F]=1,\quad [f]=[\bar f]=\frac12,\quad [f^{(-q_{1},-q_{2})}] =0\] and the parameters of the superconformal algebra $\mathfrak{su}(2,2|3)$ have mass dimensions 
\[[a]=1,\quad [q]=[\bar q] = \frac12,\quad [s]=[\bar s]=-\frac12,\quad [k]=-1\] and all others zero. The fermionic expansion of the components of the Killing vector field are uniquely determined by the above constraints, the mass dimensions and the harmonic charges. 

Since $F$ satisfies \eqref{eq:killing} there exists an expansion \cite{West:1997vm}
\begin{align}
F_{\beta\dot\beta} = a_{\beta\dot\beta}(\theta,\bar\theta) &+ b(\theta,\bar\theta) x_{\beta\dot\beta} + (\delta^{\dot\alpha}_{\dot\beta}c^{\alpha}{}_{\beta}(\theta,\bar\theta) + \delta^{\alpha}_{\beta}\bar c^{\dot\alpha}{}_{\dot\beta}(\theta,\bar\theta))x_{\alpha\dot\alpha}\notag\\&\quad + d_{\beta\dot\beta}(\theta,\bar\theta)x^{2} - 2(d(\theta,\bar\theta)\cdot x)x_{\beta\dot\beta}.
\end{align}
The coefficient superfunctions $a$ through $d$ have fermionic expansions with fixed parameters. In fact $d_{\alpha\dot\alpha}=k_{\alpha\dot\alpha}$ is restricted to be purely bosonic while the rest have fermionic expansions of varying length thus containing all the superconformal transformations of the algebra $\mathfrak{su}(2,2|3)$.

Given the Killing vector field $V$ it is now possible to find the action on the gauge connection one-form $A$. $A$ transforms like a scalar \[A'(X')=A(X)\] and its transformation is given concisely by the Lie derivative along the vector field $V$ \[\delta A = \mathcal{L}_{V}A.\]

\bibliographystyle{JHEP}
\bibliography{harmonic}

\end{document}